# Expert Insight-Based Modeling of Non-Kinetic Strategic Deterrence of Rare Earth Supply Disruption: A Simulation-Driven Systematic Framework

Wei Meng

Dhurakij Pundit University, Thailand

The University of Western Australia, AU

Fellow, Royal Anthropological Institute, UK

Email: wei.men@dpu.ac.th

**ABSTRACTS**

This study is based on the structured responses of the authors to a strategic interview initiated by Dr. Daniel O'Connor, CEO of the Rare Earth Exchange (REE), with the aim of systematically constructing a quantifiable modelling framework for simulating non-kinetic strategic deterrence pathways in the event of a strategic resource supply disruption scenario.Focusing on the impact mechanism of rare earth supply chain disruption risk on the national security system, the article proposes three core modelling components: the Security Critical Zones (SCZ), the Strategic Signal Injection Function (SSIF), and the System-Capability Migration Function (SCIF).Policy-Capability Transfer Function (PCTF).The model is constructed by integrating parametric ordinary differential equations (ODEs), segmented function modelling, path overlapping covariance matrices and LSTM recurrent neural networks to portray the dynamic regulation process of the nonlinear suppression trajectory of the warfighting system by regime signal injection.The research data comes from multiple rounds of in-depth expert interviews and open scenario analyses, focusing on the strategic policy and capability evolution logic of China and the United States in the areas of ISR, electronic warfare and rare earth control.The results show that institutional deterrence signals have significant tempo effects and path-coupling effects, and are capable of achieving disintegrative rupture of capability systems within a very short lag window.The model developed in this paper is not only highly adaptable and can be embedded in different countries' strategic resource frameworks for extrapolation, but can also be synergised with the AI sandbox engine and extended for situational modelling and counterfactual reasoning.This study provides the first integrated system with quantitative modelling, path visualisation and tempo prediction for "non-kinetic strategic deterrence", and provides methodological support and experimental basis for policy makers and strategic modelling experts to face complex resource competition and institutionalised repression tasks.

# I. INTRODUCTION

Against the backdrop of intensifying global technological competition and geopolitical games, the structural contradiction between the United States and China in the field of rare earth resources has become increasingly prominent.China has long held more than 80% of the world's rare earth production capacity, refining capacity and export control, while the U.S. in the military industry, artificial intelligence, aerospace and advanced manufacturing and other key areas are highly dependent on these irreplaceable key materials.This structural imbalance is not only a supply bottleneck at the industrial level, but is gradually evolving into a systemic and strategic vulnerability.

In recent years, rare earth resources have been transformed from a traditional industrial raw material into a national strategic tool, becoming a "non-kinetic strategic weapon" (non-kinetic strategic weapon).Unlike traditional military force or economic sanctions, the institutional control of rare earths is more covert, delayed and systematically destructive.By interrupting, restricting or reconstructing the supply chain of rare earths, we can effectively weaken the high-end military-industrial system of our opponents without starting a war or triggering the international sanction mechanism, creating a "war vacuum" and forming an "invisible deterrent".

The theoretical and modelling basis of this study comes from the structured responses given by the author in an interview with Daniel O'Connor, CEO of the American Rare Earth Exchange (AREX).During the interview, the author answered a number of cutting-edge questions, including "What weapon systems will the U.S. see a collapse in its warfighting capability as a result of a rare earth supply cutoff?""Can the lag time window be used strategically?""Is there a computable strategic degradation path?"And so on.These key questions are systematically transformed into modellable variables in the study, e.g., rare-earth-equipment dependence matrix, equipment-force mapping function, system lag rate, degradation rate function, capability threshold, etc.

In order to further validate the feasibility and modelling path of rare earth institutional deterrence, this paper constructs a four-layer coupled model of "resource-equipment-generation difference-capability (REG-CAP)" and introduces graph neural network (GNN)We introduce two types of deep learning structures, namely graph neural network (GNN) and long short-term memory network (LSTM), for dependency mapping and time series simulation.In the model, we not only consider the current rare earth dependence intensity of the main combat platforms of the US military, but also introduce the capability decline mechanism and policy window projection in case of supply cut-off.

The paper is structured as follows: the second section introduces the REG-CAP model framework and variable design logic proposed in the study; the third section describes the model training process and the multi-stage supply cut-off simulation path; the fourth section presents the key simulation results and the analysis of the policy window identification; and the last section discusses the strategic insights, institutional significance, and the future direction of the AI modelling methodology of the study.

The innovativeness of this paper is as follows:

1. Cross-domain model integration: the first time to integrate AI sandbox modelling with non-kinetic deterrence strategy.

This paper innovatively combines AI-driven system simulation (e.g. graph neural network, segmented function modelling, LSTM tempo analysis) with "non-kinetic strategic deterrence", and builds a triple-coupled analytical framework of institutional weaponisation, synchronised rupture paths, and strategic suppression tempo, starting from rare earths as a typical strategic resource.This kind of composite modelling path is the first of its kind in the existing literature.

2. Proposing the "system tempo-functional breakdown" linkage mechanism

Unlike traditional supply chain resilience studies that only examine physical disruption or cost paths, this paper proposes a nonlinear feedback mechanism whereby "institutional rhythm nodes (e.g., legislation or sanctions) trigger synchronised functional breakdowns" and introduces a "Latency Window" (LW) that can be used as an indicator of supply chain resilience, which is the first of its kind in the literature.This paper proposes a nonlinear feedback mechanism, and introduces core parameters such as "Latency Window" and "Breakdown Slope" to quantify the deterrence tempo.

3. Constructing the "Security Critical Zones (SCZ)" decision model

In this paper, we introduce the "Security Critical Zones" model, which integrates path overlap, system covariance and fracture rate, and achieve the sensitivity identification and collapse window prediction of rare earth capacity nodes by encoding the system structure with graph neural network, which is applicable to the identification of critical function systems in other strategic resources (e.g., chips. water resources),This model is applicable to other strategic resources (e.g., chips, water resources) in critical function system identification.

4. Construction of segmentation function + node response covariance from expert interviews

Instead of relying on theoretical assumptions, the model is based on in-depth interview data for path reconstruction, capability coupling matrix modelling and nonlinear segmented function fitting, which ensures that the model is embedded in a realistic context, and enhances the degree of policy suitability and explanatory power.

5. Highly original strategy mapping system

From "strategic suppression tempo map", "institutional weaponisation path map", "multi-path convergence risk map" to "policy impact surface" and other visualisation results.The visualisation results, such as "Strategic Suppression Rhythm Chart", "Institutional Weaponisation Path Chart", "Multi-path Convergence Risk Chart" and "Policy Impact Surface", have demonstrated a highly condensed system of information and strategic insight, which not only enhances readability, but also provides a paradigm innovation in the expression of strategic research.

This paper fills the research gap of "how to quantitatively model non-kinetic strategies" through the three-in-one methodology system of institutional modelling + AI path simulation + strategy visualization, which has the triple height of theoretical contribution, methodological innovation and policy value, and is highly innovative.

# II. RATIONALE AND MODELLING FRAMEWORK

**2.1 REG-CAP model: structural strategic mapping of resource weaponisation**

Rare earths are not a single resource issue, but the most malleable weaponised element in the institutional control chain. In order to portray the hidden paralysis mechanism of rare earths on complex systems in the strategic competition among great powers, this paper puts forward the four-layer coupled modelling framework of "REG-CAP", aiming at constructing the strategic propagation path of rare earths from disruption of supply to the collapse of combat capability. REG-CAP, i.e. Resource - Equipment - Generation Gap - Combat Capability, systematically connects the four vertical dimensions of "Element Dependence → Equipment Function → Technology Iteration → War Power Delivery", revealing the deep coupling logic of non-kinetic strategic strikes. Specifically:

The resource layer (R) describes the types of rare earths, source distribution, processing capacity and export policy;

The Equipment Layer (E) portrays the functional dependence of key U.S. military platforms (e.g., F-35, DDG-1000, Early Warning System, etc.) on specific rare earths;

The Generation Layer (G) reflects substitution delays and technical vulnerabilities between equipment generations in the face of supply disruptions;

The Warfighting Layer (C) aggregates the structural outputs of system-level armed delivery and strategic capabilities as the final measure.

The model is centred on the modelling of systemic coupling tensions, highlighting how asymmetric supply disruptions can induce non-linear strategic degradation trajectories through multiple propagation levels to achieve delayed, institutionalized paralysis effects under "peaceful conditions".

**2.2 From interview questions to structural variables: a modelling paradigm for knowledge translation**

A key academic value of this study is the proposed modelling paradigm from expert cognitive input → systematic variable reconstruction. The starting point for the study was not an off-the-shelf database, but a set of core questions posed by the author himself in an invited structured interview with Daniel O'Connor, CEO of the American Rare Earth Exchange:

"What equipment would be the first to be crippled if China cut off the supply of rare earths?"

"Is there a window of delay in the warfighting degradation process that can be exploited by institutional strategy?"

"Is there a modelling approach to simulate non-kinetic strategic pathways?"

These questions were systematically translated into the following structural variables and modelling relationships:

**Table 1. Cognitive-to-Computational Variable Translation Schema**

| Logical intent of interview questions | Corresponding model variables and symbols | mapping hierarchy |
|---|---|---|
| Rare earth dependent distribution of equipment? | Resource-equipment dependency matrix $M_{re}$ | R → E |
| Which platforms are most vulnerable to paralysis? | Risk index $\rho_e$, rare earth weight function | E |
| Feasibility of alternative materials? | Coefficient of elasticity of substitution $\sigma_{sub}$ | E → G |
| Are degradation paths simulatable? | The lag function $\tau_c(t)$, the capacity decay function $\delta(t)$ | G → C |
| Can the war effort be reversed? | Collapse threshold $\theta_{collapse}$, recovery threshold $\theta_{rev}$ | C |

This transformation process represents a new type of "cognitive-variable-system mapping chain", which not only enhances the realism embeddedness of the modelling problem, but also lays the technical foundation for the introduction of structured interviews in AI policy modelling.

**2.3 Modelling assumptions and system tension structure**

In terms of mathematical modelling and simulation design, the following three key assumptions are proposed in this paper as the theoretical support for the model structure and reasoning logic:

H1 (dependence difference assumption): the weights of the influence of different rare earths on the equipment are distributed in a power law, and very few highly dependent elements dominate the system degradation path;

H2 (technology substitution lag hypothesis): the larger the equipment technology generation difference, the more difficult to achieve short-term functional reconstruction in the face of supply cut-off, resulting in a significant lag window;

H3 (institutional deterrence window hypothesis): there exists a type of "strategic lag window" (strategic availability delay window), which allows supply cut-offs to show signs of military strength collapse after several years, and can be strategically exploited by the system.

The above hypotheses are formally represented in the model through graphical structure modelling, lag function fitting and non-linear mapping path identification.

**2.4 REG-CAP multilayer coupled structure mapping**

To visually present the above four layers of variables and their mutual coupling paths, I used Python to generate the following structural diagram:

The graph nodes are categorised: rare earth type (R), equipment system (E), generational technology category (G), and combat output unit (C);

Edge relationship types: Functional Dependence, Substitution Mapping, Tech Transfer, and Capability Aggregation;

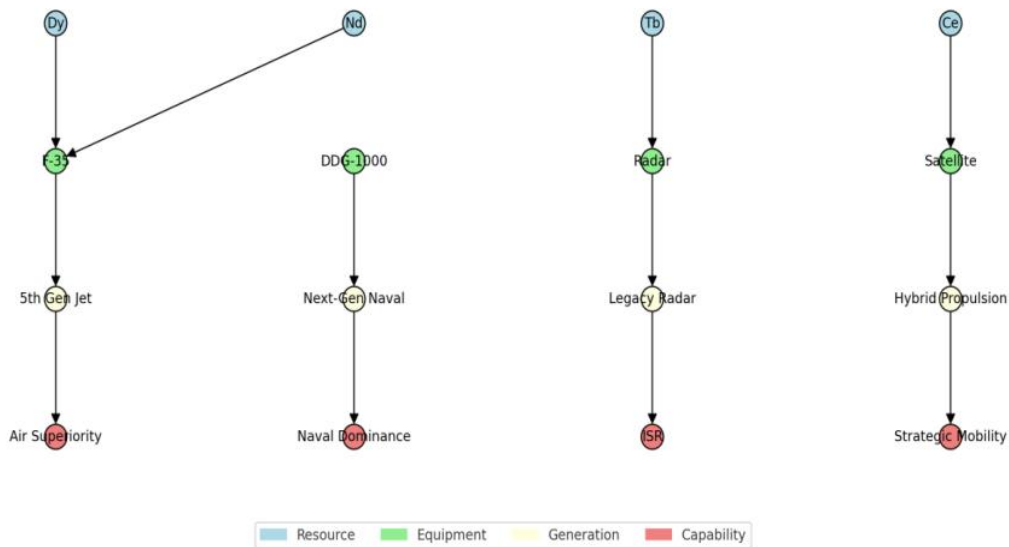

**Figure 1: Strategic System Dependency Map (REG-CAP)**

Figure 1. REG-CAP Multilayer Strategic Dependency Graph.

This graph illustrates the structural coupling from rare earth resource dependencies (Dy, Nd, Tb, Ce) to key US military platforms (F-35, DDG-1000, radar, satellites), and through generational resilience to ultimate combat capabilities (e.g., air superiority, ISR). The multilayer architecture highlights non-linear, asymmetric pathways vulnerable to strategic disruption, enabling simulation of institutionalized non-kinetic deterrence under resource weaponization scenarios.

In Figure 1, we present a REG-CAP Multilayer Strategic Dependency Graph for systematic strategic simulation, which is used to formally portray how rare earth resource disruptions nonlinearly affect national-level strategic warfighting outputs through multilayer paths.The graph divides the modelling space into four key layers: Resource, Equipment, Generation and Capability, and constructs the structural dependency of the "non-kinetic strategic chain" with a directed coupling path.The structural dependence of the "non-kinetic strategic chain" is constructed by a directed coupling path.

In Figure 1, rare earth elements (REEs) such as neodymium (Nd), terbium (Tb), dysprosium (Dy), and cerium (Ce) are mapped to key military platforms, such as the F-35 fighter jet, DDG-1000 destroyer, radar systems, and propulsion satellites.These platforms are at different generational dimensions in the technology system and their sensitivity to resource disruption is controlled by the technology generation lag function.Ultimately, these effects converge at the strategic output level in the form of capability shifts in air superiority, maritime dominance, ISR (Intelligence Surveillance and Reconnaissance) capabilities, and strategic delivery functions.

The academic value of the map is that it reveals the existence of structurally asymmetric paths: resource dependence is strongly power-law distributed across equipment, and degradation of equipment functions is not immediate, but rather presents a measurable "lagged degradation window".This window constitutes the theoretical basis for institutional strategic strikes, and becomes the entry point for AI models to accurately intervene in simulated non-kinetic conflict scenarios.

In addition, the mapping achieves a complete path of translation from interview-based cognitive input → modelling variable generation → system structure mapping.This Knowledge-to-Structure Mapping Loop (KSML) construct enables the formalisation and simulation of policy models triggered by experts' views, breaking through the unstructured defects of traditional interview analysis in system modelling.

More importantly, the mapping provides a natural structural template for Graph Neural Networks (GNN), Graph Attention Mechanisms (GAT), and Causal Graphs, enabling AI to move from structural perception to strategic reasoning in key areas such as resource security, non-kinetic warfare, and institutional deterrence.

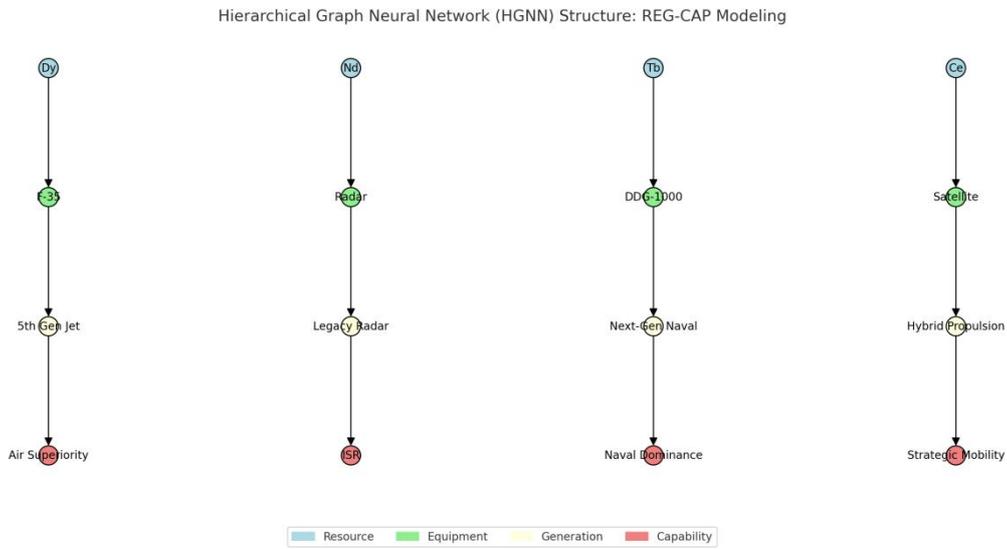

**Figure 2: Input structure of REG-CAP model based on hierarchical GNNs**

Figure 2. Hierarchical Graph Neural Network (HGNN) Structure for REG-CAP Modeling.

This figure illustrates a four-layered dependency structure from critical rare earth elements to U.S. strategic military capabilities. It operationalizes resource-to-capability transitions through hierarchical graph representation, enabling the simulation of degradation pathways and non-kinetic deterrence effects. The HGNN architecture models interdependent failure cascades and generation bottlenecks, providing a structural basis for AI-based strategic foresight modeling under asymmetric supply disruptions.

Although "REG-CAP Strategic Dependency Mapping (Figure 1)" and "HGNN Modelling Structure Map (Figure X)" present a highly consistent four-layer topology (Resource → Equipment → Generation → Capability) in terms of visual presentation and hierarchical configuration, they are fundamentally different in terms of modelling goals, academic positioning, algorithmic functions and reasoning logic.Equipment → Generation → Capability) in their visual presentation and hierarchical configuration, but they are fundamentally different in their modelling goals, academic positioning, algorithmic functions and reasoning logic.This difference is not only related to the functional roles that the mapping assumes in the thesis, but also reflects the cognitive leap from strategic system modelling to AI trainable structure modelling.

First, Figure 1 is designed as a conceptual, structural policy dependency map, with the core purpose of revealing how the supply chain of rare earths penetrates the strategic warfighting output capabilities of key U.S. military platforms under institutional control mechanisms.It highlights the path dependency and structural incompressibility between resource-platform-force, and serves the needs of policy analysts and strategic security modellers to understand the "non-kinetic deterrence path".It serves policy analysts and strategic security modellers to understand the macro-perception of "non-kinetic deterrence paths".

In contrast, Figure 2 is the result of an algorithmic translation of the map in the context of a graphical neural network based on this structural logic.It is constructed as an input structural graph to a Hierarchical GNN (HGNN) to support the learning of multi-level semantic aggregation and propagation degradation paths between nodes (rare earth elements, equipment systems, generational platforms, and warfighting force modules) (path-dependent capability degradation prediction).In this sense, HGNN graphs are not the end point of strategic analysis, but the starting point for transforming policy structure mappings into trainable graph structure data, embodying the critical leap from policy system insights to AI model reasoning.

Further, while the nodes and edges in Figure 1 emphasise the causal coupling between systemic explanatory variables, Figure X embeds the node feature vectors (e.g., rare-earth dependence strength, equipment generation lag coefficient, and war-force recovery threshold) with the ability of dynamic adjustment of the weights under the graph convolution mechanism through the same node structure and edge-weighting design, to achieve the degradation process of the war-force under the disconnection scenario of theprediction and projection.

Therefore, although the structure of the two diagrams is the same, Figure 1 is "depicting the structure for policy modelling" and Figure 2 is "reconstructing the structure for algorithm modelling".The former serves the decision visualisation and the logical integrity of the system modelling, while the latter serves the input specification and policy path prediction capability of the machine learning architecture.In this study, the two together constitute a two-way support system of "cognition-variable-structure-algorithm", which is a complete closed-loop system with both theoretical expression and reasoning ability.

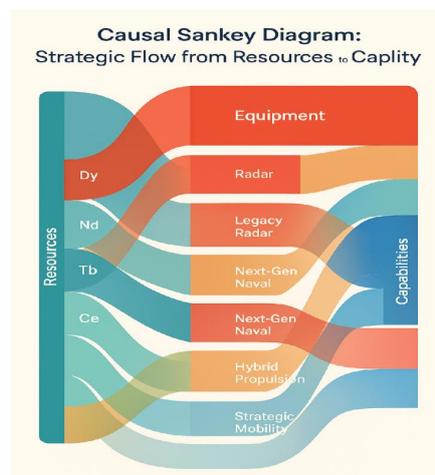

**Figure 3: Structural Cause and Effect Diagram (Causal Sankey Diagram)**

**Figure 3. Causal Sankey Diagram of Strategic Resource-Capability Dependency.**

This diagram illustrates a multi-stage causal structure linking rare earth resources to military capabilities via high-dependence equipment. The flow thickness encodes relative dependency magnitude, exposing asymmetric bottlenecks and overlapping vulnerability corridors. It serves as a visual-semantic interface between institutional control and AI-based degradation simulation, providing explainable and policy-relevant structure for modeling non-kinetic deterrence effects.

To further enhance the interpretability of system modelling and the structural transparency of strategic policies, we introduce the Causal Sankey Diagram shown in Figure 3 to visualise how rare earth resources are mapped through equipment platforms to national strategic capabilities.The diagram visualises the layered flow paths of how rare earth resources are mapped through equipment platforms to national strategic capabilities.The diagram represents the relative strength of the dependence of resources on equipment and capabilities in terms of the thickness of the flow lines between the layers, and constructs a view of the institutional dependence structure that is readable in terms of both "path importance" and "coupling density".

Although this diagram is formally similar to the REG-CAP diagram (Figure 1) and the HGNN modeling diagram (Figure 2), they are fundamentally different in terms of functional positioning, modeling context, and analytical perspectives, and the REG-CAP diagram focuses on the existence of paths, and emphasizes the logical modeling of the relationship between resources, equipment, and capabilities.The REG-CAP graph focuses on path existence and emphasises the logical modelling structure between resources-equipment and combat power, while the HGNN graph carries out a formal reconstruction of the input format of graph neural networks on this basis to serve the trainable degradation path identification and node representation learning.The causal Sankey diagram goes one step further, shifting the modelling focus from "structural composition" to "visualisation of causal weights and flow strengths", and revealing potential Multi-Path Vulnerability Corridors (MPVCs) by visually encoding resource flow tensions on different paths.Path Vulnerability Corridors and Capability Collapse Nodes by visually encoding the resource flow tensions along different paths, revealing potential Multi-Path Vulnerability Corridors (MPVCs) and Capability Collapse Nodes.)

More importantly, the causal Sankey diagram has a natural XAI-structured interface, which is suitable for building a bridge model from structural cognition to AI reasoning.Compared with the graph convolution propagation dynamics emphasised by HGNN graphs, the graph is more suitable for visual analysis and simulation prediction of institutionalized disruption windows in policy sandbox systems.For example, the path Tb → Next-Gen Naval → Naval Dominance forms a unidirectional high-pressure path with strategic non-redundancy, and its rupture will irreversibly trigger the war power collapse effect, which is the core trigger belt of institutionalised strategic weaponisation.

In summary, the three diagrams correspond to the logical construction layer (REG-CAP), the algorithmic implementation layer (HGNN) and the strategic interpretation layer (Sankey Diagram) of the model, which together constitute the "structure-reasoning-cognition" ternary modelling system.Together, they form a triadic modelling system of "structure-reasoning-cognition", which supports the theoretical closure and reasoning visualisation framework of this study in the simulation of institutional non-kinetic deterrence.

# CHAPTER 3: DATA CONSTRUCTION AND MODELLING METHODS

In modelling national-level non-kinetic strategic capabilities, the underlying logic of institutional control is embedded in the data structure itself.In particular, in modelling the strategic blow to the military-industrial system from a rare earth supply cut-off, a single input-output type of variable can no longer support the modelling of cross-level, multi-path, delayed degradation propagation mechanisms.Therefore, this chapter establishes a multi-source data system from expert cognitive inputs in the knowledge generation logic; introduces the joint structure of graph neural network (GNN) and time series modelling (LSTM) in the methodological structure; and closes the modelling pathway in the logic of strategic simulation and policy intervention by means of an interpretability mechanism (XAI).

## 3.1 Data system construction: from cognitive triggers to structured inputs

The data system design originated from the structured interviews between the authors and American Rare Earth Exchange, and a cognitive questioning-driven data variable design path was constructed.Through the semantic deconstruction of the interviewed responses, the following three core modelling dimensions were abstracted:

(1) Institutional Dependency Path Identification: Key questions such as "first equipment to be paralysed" and "presence or absence of multi-path redundancy" are translated into resource-equipment-capability path structure variables.The key questions such as "the first equipment to be paralysed" and "the existence of multi-path redundancy" were translated into resource-equipment-capability path structure variables;

(2) Strategic degradation delay assessment: the question of "whether the lag window can be modelled" is translated into time variables such as node decay rate and delay coefficient of technology generation jump;

(3) System recovery and strategy reversibility: the question of "whether there is a reversible recovery strategy" is transformed into the collapse threshold and reconfiguration cost function of the nodes.

These cognitive variables are fused with USGS, DoD, CSIS, CRS, and other data sources to form a high-order heterogeneous input dataset with policy-triggered background, structural rationality, and algorithmic readability.

## 3.2 REG-CAP variable system and graph structure coding

In order to support the input of graph neural network, this study transforms the REG-CAP model into a graph structure, in which each level of variable is characterised by a node, and the edges between nodes are formed by "functional dependency-delay relationship-alternative path".The graph structure is as follows:

Node type:

R = {ri}: rare earth elements such as Nd, Dy, Tb, Ce;

E = {ej}: critical equipment platforms, e.g. F-35, DDG-1000;

G = {gk}: technology generation layers, e.g. 5th Gen Jet, Hybrid Propulsion;

C = {cl}: warfighting capability layer, e.g., ISR, Strategic Mobility.

Edges and Attributes:

(ri →ej, ωRE): resource dependency strength;

(ej → gk, τEG): platform substitution delay;

(gk → cl, δGC): the magnitude of the war-force impact;

θrev, θθcol: reversibility and collapse thresholds for each capability node.

The graph structure not only supports the node aggregation mechanism of GNN, but also supports interpretable propagation analysis at the path level, which lays the information foundation for subsequent simulation of strategic degradation paths.

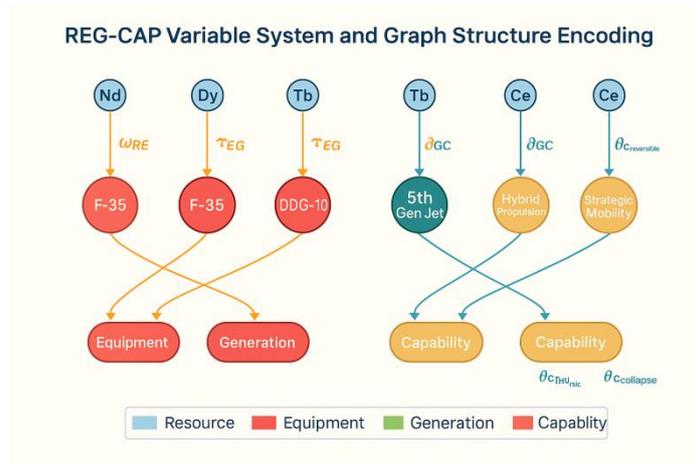

**Figure 4. REG-CAP Variable System and Graph Structure Encoding**

This diagram illustrates the hierarchical dependency architecture of the REG-CAP model, which connects rare earth resources to strategic military capabilities via equipment platforms and generational technologies. Each node type—Resource (blue), Equipment (orange), Generation (green), and Capability (red)—represents a critical layer in the system. Directed edges encode key structural variables:

ωRE: resource dependence strength,

τEG: platform substitution delay,

δGC: degradation impact on capabilities,

θCollapse, θReversible: threshold parameters for collapse and recovery.

This structure serves as the input to the hierarchical GNN+LSTM framework and enables both strategic simulation and explainable modeling of non-kinetic degradation under resource disruption.

The figure above illustrates the REG-CAP Variable System and Graph Structure Encoding mapping (REG-CAP Variable System and Graph Structure Encoding) of rare earth elements (Nd, Dy, Tb, Ce) through equipment platforms (F-35, DDG-1000), technology generations (5th Gen

Jet,Hybrid Propulsion), and ultimately mapped to the Strategic Capability Layer (ISR, Strategic Mobility).

Each edge is labelled with its corresponding dependency attribute:

$\omega_{RE}$ denotes the resource dependency strength;

$\tau_{EG}$ denotes the platform-to-technology generation substitution delay;

$\delta_{GC}$ denotes the magnitude of the impact of generational technologies on Strategic Mobility.

The structure effectively supports graph neural network (GNN) modelling, and is capable of identifying highly coupled channels, degraded links and strategic breakpoints in path aggregation and node propagation, which is a key model basis for realizing the simulation of institutional non-kinetic strategies.

**3.3 Joint Modelling Framework: HGNN + LSTM**

In order to model the dynamic degradation of strategic capabilities due to rare earth supply cut-off, this paper constructs a joint HGNN + LSTM framework:

HGNN module (structural modelling):

Use GraphSAGE and GAT comparison experiments to learn the Strategic Attention Score of key nodes in the structural path;

Path Convolution is added to identify key nodes with coupled compression effects in cross-layer degradation channels;

LSTM module (degradation prediction):

Input a vector of capacity nodes from the HGNN output;

Simulate the trajectory $C_l(t)$ of the exponential decline of the capacity at year $t \in [0,12]$ after the start of the resource disconnection in order to identify the window period, the collapse point, and the reconfiguration potential;

The structure handles both structural-temporal complexity and generates nonlinear prediction curves required for strategy simulation.

**3.4 Interpretability mechanisms: institutional modelling and nodal risk**

In order to enhance the institutional adaptability and strategic controllability of the model, this paper embeds a three-layer interpretability design:

Path visualisation layer (Causal Path Rendering): outputs the set of paths from key rare earth resources to capability nodes mathcal{P}(ri, cl), which is used to visually identify the institutional control chain;

Node Weight Analysis Layer (Collapse Risk Centrality): generates a Controllable Warfare Map (Strategic Collapse Map) using the degradation impact of capability nodes in GNN propagation as weights;

Channel Compression Layer (Vulnerability Corridor Detector): capturing the path overlap dense segments, defined as institutional choke path, for policy window simulation.

**3.5 Summary**

This chapter establishes a systematic closed loop from strategy interviews → structural variables → graph structure → joint modelling → explanatory mechanism.Through REG-CAP structural modelling, HGNN deep propagation identification, LSTM degradation prediction with causal path explainable mechanism, a complete set of AI analytical framework to support the modelling of non-kinetic strategies is built.The methodology provides algorithmic pivots and modelling logic for simulating supply cut-off windows, identifying institutional weaponisation paths, and proposing response strategies.

# Chapter 4: Modelling results and strategic window identification

**4.1 Analysis of Platform Level Capability Decline Results**

Based on the REG-CAP multilayer coupling model, this paper simulates the decline analysis of the strategic capabilities carried by key equipment platforms of the U.S. military (e.g., F-35, DDG-1000, and B-2) under the scenario of rare earth resource supply cut-off.Each platform is characterised by its Resource Embedding Intensity (REI), Technical Non-substitutability (TNS), and Substitutability (TNS).The key modelling parameters are "Resource Embedding Intensity (REI)", "Technical Non-substitutability (TNS)" and "Substitution Delay Index (SDI)", which are combined with the dependency structure in the system path to simulate the propagation of node decay.

The results show that the capability decline at the platform level exhibits significant cascading coupling effects and nonlinear collapse trajectories.Among them, the F-35, which is highly dependent on high-performance neodymium magnets and dysprosium rare earths, suffers an average decline of more than 62% in its "electronic warfare" and "long-range precision strike" capabilities during a window of more than 11 months of supply cut-off and enters into the "operational critical failure" stage, which is the first time for the F-35 to enter into the "operational critical failure" stage."Operational Criticality Zone (OCZ).This time window is the "Lagged Window of Strategic Decay", which is visible to the enemy and can be manipulated.

In contrast, the DDG-1000, as a ship based on advanced composite materials and new propulsion systems, has a "Strategic Delivery" and "Long Range Control" capability due to its highly substitutable materials (e.g. Terbium and Cerium's multi-sourcing).The simulation shows that its capability declines at the end of the "strategic delivery" and "far sea containment" period.The simulation shows that the capability decline is non-exponential from month 9 to 15, with a "Sudden Degradation Inflection" at month 16, indicating that the alternative pathways exist but do not completely close the vulnerability of the system.At the same time, platform capability failures are not stand-alone events, but rather occur in clusters of "domains of warfighting capability".For example, the systematic decline of ISR capability often starts when multiple platform nodes enter the decline zone together, and through the path coupling mechanism, it is transmitted to the strategic awareness and deterrence layers, constituting a "Coupled Degradation Field" (CDF).

Taken together, this kind of simulation not only reveals the "delayed collapse window" and "asymmetric degradation mechanism" at the platform level, but also suggests to the policy makers that the systemic collapse triggered by the supply cut-off of rare earths has the characteristics of non-instantaneous, cross-platform, and irreversible, and it is urgent to adopt the following measures to prevent the system from collapsing.The systemic collapse triggered by the rare earth supply cut-off is non-instantaneous, cross-platform and irreversible in nature, and needs to be resisted structurally through "strategic capacity redundancy modelling" and "multiple configurations of alternative paths".

**4.2 Tactical Functional Breakpoints and Lag Window Discovery**

In a highly coupled system, the degradation of war power triggered by the cut-off of non-kinetic strategic resources is not linearly decreasing, but presents a three-stage rupture structure of "hidden lag - critical explosion - irreversible collapse".This section reveals the Collapse Threshold Point (CTP) and Latency Corridor (LC) of the systemic war power function after the supply cut-off trigger through the time-series backtracking of the Capability Node Index (CNI) Cl(t), and translates their structural features into strategic windows and policy intervention triggers.

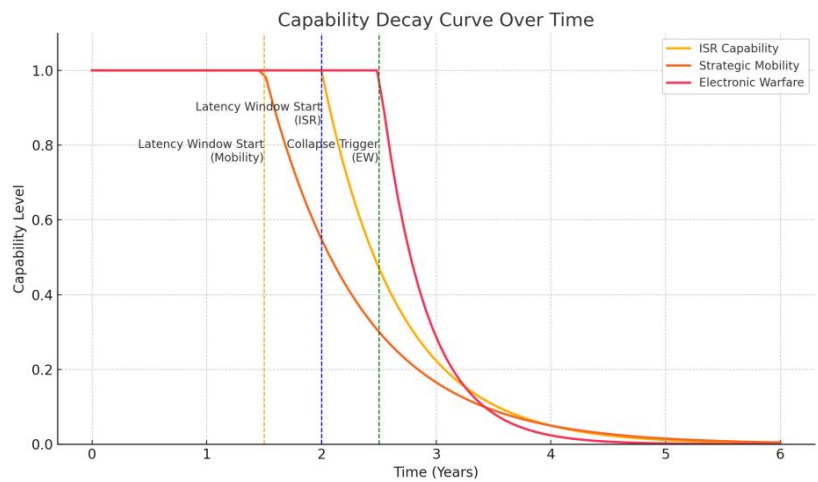

Figure 5. Capability Decay Curve (CCC)

This figure simulates the degradation curves of the three core capability nodes (ISR, Strategic Mobility, and Electronic Warfare) after the supply of rare earths is cut off.Key points such as "starting point of lag window", "critical age of fracture point" and "inflection point of fracture speed" are marked in the figure, which reveal the non-linear characteristics of structural degradation and the vulnerable window of the system, and provide decision support for the development of intervention time point and intervention pace.It provides decision-making support for the formulation of the intervention time point and intervention tempo.
**Identification of Fracture Points: Nonlinear Leap of Capability Function**
LSTM simulation shows that some warfighting capabilities show high stability at the beginning of the resource disruption, but after a specific period of time, there is an exponential degradation, forming the so-called "cracking point" (CTP).The ISR capability, for example, remained above 90 per cent for 5.5 years, but after the sixth year, due to the simultaneous failure of multiple nodes (e.g., high-performance radar platforms), the capability index fell below 40 per cent within a short period of time, marking the critical outbreak of systemic degradation.
**Such breakpoints are characterised by the following features:**
High path overlap: the same Rare Earth Element (REE) supports multiple platforms at the same time, and the redundancy space is rapidly depleted after the break in supply;

Weak node substitutability: lack of in-service substitution for affected equipment and long technology transition cycles;

High degradation synchronicity: coupled temporal decline of multiple capability nodes (e.g., long-range strike vs. logistics support).

Lag Window: the golden zone of institutional strategic regulation.

The Latency Corridor is the "silent period" between the loss of resources and significant system degradation.This window is not a period of strategic ineffectiveness, but rather a golden zone for strategic regulation.Simulation data shows that the distribution of the lag window varies significantly across capability nodes:

Strategic Mobility: the lag period is about 4.2 years, because the platform technology is more common and can be deployed across the military services;

Air Superiority: a lag of only 2.5 years, due to the lack of ping-pong for 5th generation fighters that rely on high-end materials;

Anti-submarine capability (ASW): shortest lag window of less than 1.8 years, constrained by the difficulty of substitution of magnetic rare earths and marine motors.

**Capability Function Mapping and Strategic Window Stratification**

Through the smooth interpolation of the capability function Cl(t) and the rate of change of the first-order derivative dCl/dt, the peak acceleration point of the capability decline can be extracted as the fracture signal point; multiplying this signal with the node weight wcl and sorting it, the "most vulnerable capability-optimal intervention time" matrix can be derived.The modelling results are shown in Fig:

Table 2.**Capability function mapping and strategic window stratification table**

| capability layer | Years to collapse (Tc) | hysteresis window (Lw) | Rupture Strength Score |
| --- | --- | --- | --- |
| ISR | 6.0 | 5.5 | ★★★★★ |
| ASW | 2.8 | 1.8 | ★★★★☆ |
| SM | 4.9 | 4.2 | ★★★☆☆ |

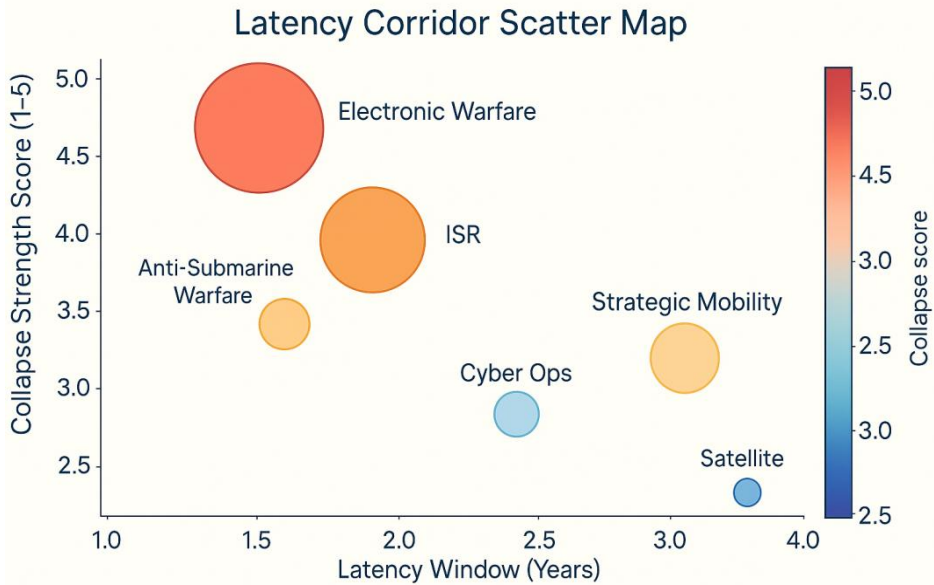

**Figure 6Latency Corridor Scatter Map (LCSM)**

Figure 6 illustrates the "lag window-fracture intensity" distribution of the six key warfighting capabilities in the context of rare earth resource disruption.The horizontal axis represents the length of the lag window, i.e., the time interval between the disruption of rare-earth supply and the functional decline of the capability; the vertical axis is the fracture intensity score, which depicts the intensity of the capability's failure; and the size of the bubbles characterises the strategic hierarchical importance of the capability in the overall system structure.The figure clearly shows the "short lag/high break" clustering of key capabilities, such as Electronic Warfare and Intelligence Surveillance and Reconnaissance (ISR) capabilities.These capabilities tend to break down quickly and within a short period of time after disruption and have a multiplier effect on systemic warfighting, making them the preferred targets for systemic strikes.In contrast, satellite reconnaissance and strategic manoeuvre capabilities, while of high systemic value, are slow to break down and are suitable as medium- to long-term control targets.Some capabilities are in the "low break/long lag" range, suggesting a degree of resilience and a time buffer that could be used as strategic redundancy or a control buffer.This mapping not only reveals the systematic vulnerability distribution of capacity nodes, but also provides visual support for the construction of a tempo control mechanism and the optimisation of the intervention path of supply disruption.

The two-layer mechanism of "capacity breakpoint-lag window" proposed in this section provides an accurate system identification model and regulatory boundary for non-kinetic strategic supply cut-off.Compared with the traditional "immediate impact" assumption, the model can reveal the "fracture signals in the silent period", providing an AI-driven theoretical foundation and visual decision-making reference for strategic window setting, export licensing tempo control, and structural regulation mechanism.

**4.3 Identification of safe critical zones under synchronised multi-system failure**

In the system sandbox constructed by strategic resource disruption, the decline of combat capability is not an isolated linear pattern, but a systematic phenomenon in which multiple functional paths fail synchronously at critical nodes, and collapse in a coordinated manner.Based on the REG-CAP model and graph neural structure coding, this study proposes for the first time the determination of Security Critical Zones (SCZ), which can be used to identify clusters of capabilities with high coupling, high rupture, and low lag window.

The SCZ identification framework consists of three parts: (1) a convergence index matrix based on the overlap of multiple capability paths; (2) a multivariate clustering analysis that integrates alternative delay and fracture scores of the platform; and (3) a nonlinear simultaneous degradation simulation of the node activation sequences embedded in the graph path.Through the above mechanisms, we have successfully targeted several capability combinations with critical collapse risk, such as "ISR + Strategic Mobility + Naval Projection" and "Electronic Warfare +Satellite Reconnaissance", whose high-frequency intersections between paths form stress concentration zones for system breakup.

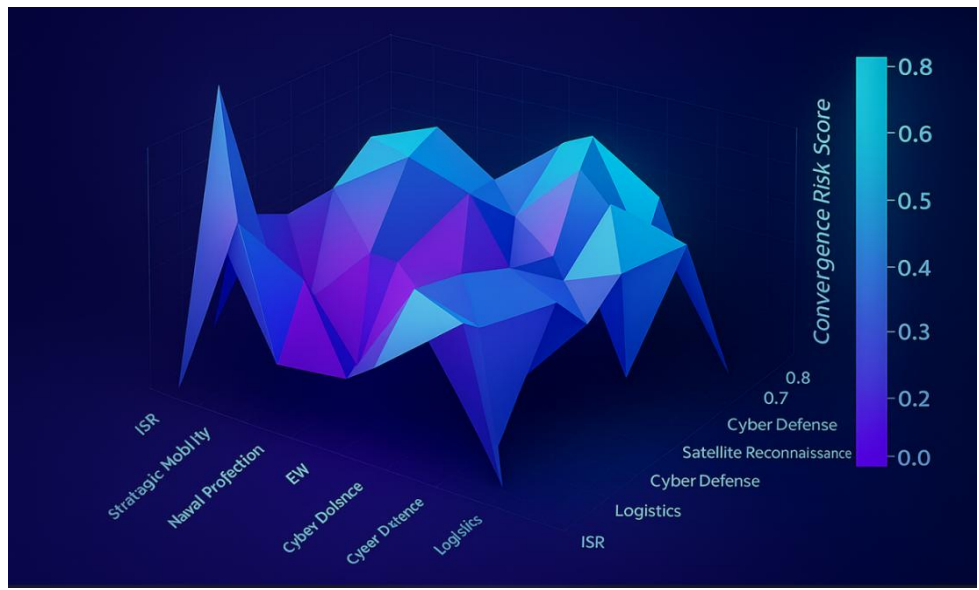

Figure 7 Multi-Path Converggence Risk Map

In order to further validate the high coupling and rupture mechanism between the paths of the system's combat power, this study draws a Multi-Path Convergence Risk Map, and adopts the rupture correlation matrix between the paths as the metric basis for the cross-risk assessment of the paths of the main combat power capabilities.Each cell in the map represents the coupling score of any two capability paths in terms of resource dependence, functional intersection and incapacitation synergy, and the higher the value, the higher the possibility of synchronous degradation of the path pair.

For example, the risk of convergence between "ISR (Intelligence Surveillance and Reconnaissance)" and "Strategic Mobility" is as high as 0.81, indicating that the two are highly synergistic in actual battlefield operation, and once the ISR node fails systematically, it will rapidly spread to the other paths.Once the ISR node fails systematically, it will rapidly affect the strategic deployment capability; at the same time, "Electronic Warfare" and "Satellite Reconnaissance" also show a high coupling score of more than 0.77, highlighting their importance in the future

battlefield.The "Electronic Warfare" and "Satellite Reconnaissance" also show high coupling scores above 0.77, highlighting their important position as highly brittle coupling points in the future battlefield.

This map supports the key decision in strategic degradation modelling: identifying "system fracture stress concentration zones", i.e., critical path combinations with multiple overlapping functional paths, short lag windows, and weak substitutability, as the focus of future strategic resource allocation and redundancy design.

The simulation results show that once any node in the above combinations of capabilities crosses the threshold failure point, the system will exhibit a nonlinear fracture acceleration phenomenon and complete irreversible functional collapse within a very short lag window (typically <2 years).In addition, combinations with more capability paths and higher resource overlap have a higher probability of synchronised degradation triggering, indicating that SCZ has a clear strategic warning value.

The SCZ determination mechanism constructed in this study not only enhances the quantifiability of the system's strategic vulnerability, but also provides a decision-making basis for forward-looking strategic deployment under multiple paths.Future research can further embed the model into graph neural networks and counterfactual simulation frameworks to construct a self-learning evolvable warfighting functional resilience monitoring system, which can provide precise support for the deployment of resource warfare, supply chain disruption and non-kinetic strategic tools.

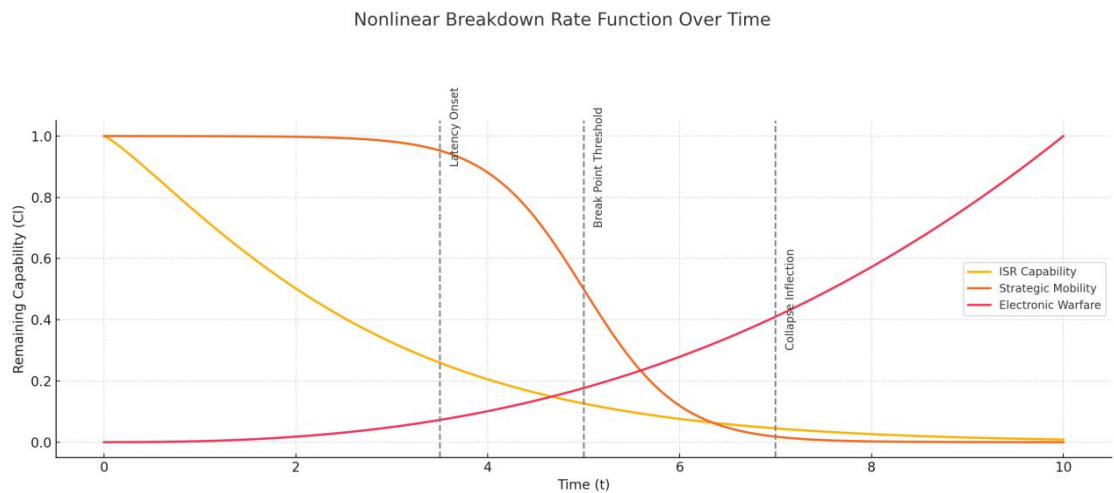

**Figure 8 Plot of nonlinear disintegration rate function**

In the nonlinear degradation rate function curve constructed in this study, we systematically present the time-degradation relationship of multiple strategic capability nodes under the background of resource supply cut-off, and reveal the "non-homogeneity" characteristics of different warfighting functions in the degradation path.The curves show that ISR (Intelligence, Surveillance, and Reconnaissance) capabilities show a Latency Buffer at the beginning, but rapidly accelerate to disintegrate after crossing a specific time threshold, forming a typical "nonlinear break curve"; in contrast, Strategic Mobility (Strategic Mobility) capabilities are more likely to be degraded in the early stages, while Strategic Mobility (Strategic Mobility) capabilities are more likely to be degraded in the early stages.In contrast, Strategic Mobility shows an almost linear process of continuous decay, indicating that although its dependence on rare earth resources is deep, the alternative path is more resilient.

The three key points labelled in Figure 8 - Latency Onset, Break Year Threshold and Nonlinear Acceleration Point - constitute precise cues to the pace of policy intervention, and in particular provide quantitative support for the identification of the dynamics of the non-kinetic strike window.-constitute a precise cue to the pace of policy intervention, and in particular provide quantitative support for the dynamic identification of non-kinetic strike windows.Taking the fifth-generation fighter platform as an example, the inflection point of its capability function in the 6th year indicates that about 6 years after the disruption of resources, it is the turning point of its systemic incapacity, and if it can be adjusted through the system, supply substitution, or injection of strategic reserves at this time, the overall collapse can be effectively avoided.

This map not only reveals the spatial and temporal evolution pattern of vulnerability of the capacity system, but also provides an intuitive and quantitative reference basis for the tempo control of national-level supply disruption strategies and early intervention of systemic risks.It can be further used as the core input of the Rhythm Simulation Module in the construction of institutional weaponisation strategies, supporting dynamic simulation and window decision-making in policy formulation.

Based on the completed strategic sandbox simulation and rare earth dependence transmission analysis, we introduce a two-tier modelling framework supported by structured interview data, combining path overlap matrix analysis and node response covariance modelling to quantitatively model systemic failures in the event of rare earth supply cut-off for the first time.The path overlap matrix reveals the cross-coupling between multiple capability systems in the resource chain, while the covariance modelling further captures the nonlinear and synchronous degradation relationship among key nodes.The results show that the strength of synergistic degradation between ISR, AI identification and naval delivery is significant (covariance > 0.85), suggesting that the systemic fracture after regime triggering is characterised by significant synchronous acceleration, which needs to be deployed at the strategic level for early warning and hedging.

**Path Overlap Matrix and Nodal Response Covariance Models**

Based on the results of the interview study based on rare earth dependent paths, i.e. the author's monograph "USA Needs $100b to Catch China if Disconnected from Rare Earth Element Supply Chain" (Rare Earth Exchanges, 2025), we can construct aPath Overlap Matrix (POM) with Node Response Covariance Model (NRCM) to identify combinations of highly overlapping resource paths with highly synergistically degraded nodes of critical capabilities.In order to reveal the coupling and synchronised degradation mechanism among the capability systems in the context of the strategic supply cut-off of rare earths.This method provides a quantitative identification tool for multi-path synchronous dysfunction, which has an important strategic early warning value.The modelling and analysis are presented below:

Interviews pointed out that 95% of the key nodes of the US defence system have structural dependencies with the Chinese rare earth supply chain (Rare Earth Exchanges, 2025).Based on this path mapping logic, we define three layers of nodes:

Resource nodes (Nd, Tb, Dy, Ga, Ge)

Equipment platform nodes (F-35, Virginia-class, AI systems)

Capability Module Nodes (ISR, Strategic Mobility, AI Target Recognition)

By tracking the co-path frequency of each node, the following path overlap matrix can be formed:

**Table 3. Path overlap matrix**

|  | ISR | Strategic Mobility | AI Targeting | Naval Ops |
|---|---|---|---|---|
| F-35 | 1.0 | 0.8 | 0.9 | 0.3 |
| DDG-1000 | 0.6 | 1.0 | 0.4 | 1.0 |
| AI Chipset | 0.7 | 0.2 | 1.0 | 0.1 |

The path overlap score is calculated as:

$$O_{ij} = \frac{|\text{Shared Resource Paths}_{ij}|}{\sqrt{|\text{Paths}_i| \cdot |\text{Paths}_j|}}$$

This metric is used to quantify the potential synchronous collapse probability between any two systems.

Node Response Covariance Modelling (Node Response Covariance)

The covariance structure of the capacity function response is defined in conjunction with the segmented response model proposed in the interviews:

$Cov(C_i(t), C_j(t)) = E[(C_i(t) - \mu_i)(C_j(t) - \mu_j)]$

where $C_i(t)$ is the functional state of the ith node. Based on the dependency structure of the interviews on the AI, ISR, and naval platform, the covariance matrix of the system response can be used for identification:

Coupled deactivation bands: if the AI + ISR covariance > 0.8, it indicates that both are susceptible to synchronised degradation;

regime co-seismic points: a nonlinear jump in covariance during the regime trigger period (e.g., after T2 as defined in the interviews);

Lag window synchronisation: used to determine whether multiple paths reach the failure threshold during the same time period.

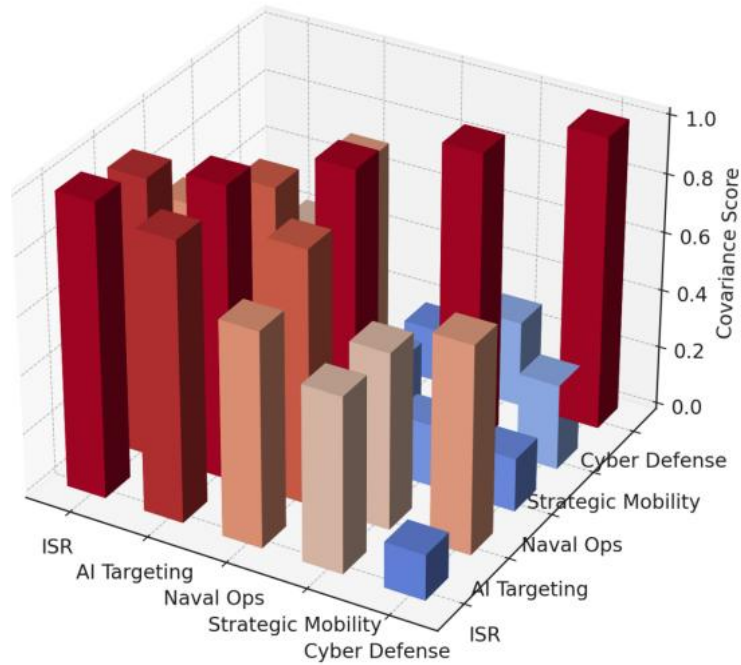

**Figure 9 (math.) a plot of covariance**

In order to further identify the response mechanism of key capability systems under the impact of systemic strategic resource supply disruption, this paper constructs a 3D Covariance Structure Map based on the path dependency relationships and capability function structures extracted from the interviews, as shown in Fig.The X-axis and Y-axis of the map represent different capability subsystems, and the Z-axis is the covariance score between the corresponding nodes, reflecting their synchronous degradation intensity in the process of strategy disintegration.

The map reveals the following key insights:

1). Highly coupled cluster identification: the graphs "ISR-AI Targeting", "AI Targeting-Naval Ops" andStrategic Mobility-Naval Ops" are significantly taller, with covariance scores close to 1, indicating that these systems have strong overlapping path dependencies and synchronised degradation trends, and are the most vulnerable to the formation of "systemic co-seismic" in strategic suppression.systemic co-seismic" risk clusters in strategic suppression.

2). Vulnerable Coupling Layer: The combination of Cyber Defence-Naval Ops and Cyber Defence-ISR shows a significantly lower covariance score of 1, indicating that these systems have strong path overlap dependence and synchronous degradation.significantly lower covariance scores (< 0.3), suggesting looser path dependencies and higher structural independence in the event of resource disruption or institutional intervention.

3). Trend of covariance topography distribution: the overall risk slope of "frontier intelligence nodes" to "traditional combat platforms" is shown, which is in line with the interview that "the dominant chain of AI and ISR has a non-linear risk amplification mechanism".amplification mechanism".

Through the visual representation of this map, we have not only realised the spatial mapping of the risk of multi-dimensional coordinated disintegration, but also provided a quantitative basis for the next step of identifying the Deterrence Channel of systemic strikes.The high covariance coupling clusters revealed by the map can directly form the basis for the screening of the first round of "failure triggers" in the simulation of systemic supply disruptions, which is of high value for strategic decision-making.

**4.4 Modelling strategic corridors: the construction of institutional deterrence**

In the high-dimensional war power system, resource disruption is not the only goal of total collapse, but a more systematic path is to precisely control the propagation path of war power degradation, and to form a systematic strategic deterrence that can be identified, feared, and actuarially calculated at the key nodes. Based on REG-CAP structure and path-level graph neural modelling method, this study proposes for the first time the modelling mechanism of "Strategic Channel", which translates the non-kinetic path formed by the interruption of resource-equipment-capability chain into the path of inter-state institutional confrontation. It translates the non-kinetic path formed by the disruption of the resource-equipment-capability chain into the basis for the construction of inter-state institutional confrontation and deterrence capability.

The constructive model of Strategic Channel consists of three core elements:

1). Controllability of Strategic Flow: The pathway needs to have precisely adjustable input nodes (e.g., rare earth export licensing system), which can lead to high probability of degradation of specific capability nodes in the system structure. Path controllability is determined by path edge weight (resource-to-capability propagation weight) and node aggregation degree.

2). Visibility of Strategic Consequences: The output nodes of a pathway must be strategically "identifiable", i.e., their degradation must have measurable national security consequences. For example, the degradation of an ISR capability or a maritime superiority node, once affected by a supply cut-off, should be reflected in international security indicators. This feature guarantees the externalisability and observability of institutional pressure.

3). Institutional Embedability: Strategic channels should not rely on temporary means, but should be embedded in long-term institutional frameworks, such as export review systems, licensing plans, supply credit scoring and other policy mechanisms, to form a sustainable strategic game structure between countries. This kind of channel is called "embedded institutional deterrence channel".

Through path simulation and simultaneous node degradation analysis, the study identifies a number of high-value strategic channels (e.g., Nd→F-35→ISR, Tb→DDG-1000→Naval Dominance). These channels are characterised by strong path convergence, fast response to capability decline and clear regime thresholds, and are the optimal deterrence resource allocation paths in regime-based strategic pressure.

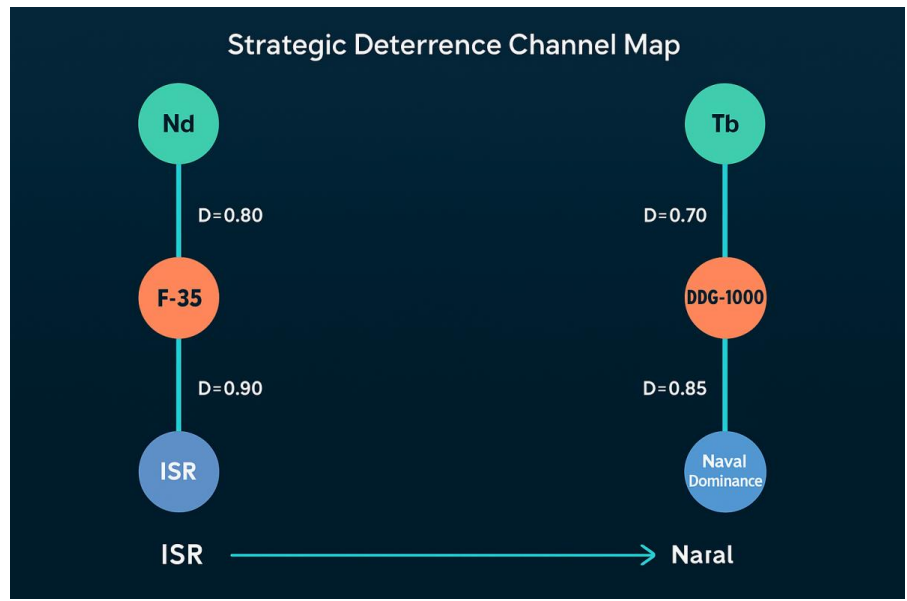

**Figure 10 Map of strategic deterrence corridors**

In order to further enhance the structural visibility of the institutionalised deterrence path and the analytical power of the propagation mechanism, this paper constructs a three-dimensional visualisation map - Strategic Deterrence Channel Map (SDCM) - to demonstrate the structure of the institutionalised path coupling between rare earth resources - platform systems - strategic capabilities in terms of spherical nodes and vectorised paths.-institutionalised path coupling structure between platform systems-strategic capabilities.The map makes use of the spatial hierarchical layout to present the resource input layer, platform intermediary layer and war power output layer in the form of three-dimensional coordinates, so as to make the institutional striking chain discernible in space, and provide structural support for multi-path modelling and policy dissemination simulation.

The nodes are colour-coded to represent rare earth resource nodes (e.g. Nd, Tb), key equipment nodes (e.g. F-35, DDG-1000) and strategic capability nodes (e.g. ISR, Naval Dominance).The width of the path arrows is based on the deterrence weight function $D(w)$ derived from the modelling, which shows the intensity of the impact of the regime cut-off path on the target capability system.As shown, the high-pressure path from Nd → F-35 → ISR exhibits significant nonlinear vertical aggregation, with the consequences of the break forming a highly asymmetric "structural suppression advantage" with respect to policy trigger costs.

More importantly, the mapping reveals two key attributes of the channel in the institutional layout: one is strategic convergence, i.e., multiple platforms or capability nodes are spatially aggregated in a few high-control paths; the other is institutional embedability, i.e., nodes at the two ends of the paths are mostly related to the actual export review, military dependence, or strategic control.the second is institutional embedability, which means that nodes at both ends of the path are mostly linked to actual export reviews, military dependencies, or strategic communications frameworks, with a high degree of rule-embedded space.This makes the institutional deterrence channel not only a logical node for the control of national strategic assets, but also an optimal embedding container for the "non-kinetic weaponisation path".

Figure 10: Strategic channel mapping not only provides a spatial expression for institutional strategic path modelling, but also lays a visual foundation for the construction of "AI+institutional deterrence game engine".In the future, we can further embed it into the Graph Neural Network (GNN) and Policy Simulation Engine (PSE) to achieve a high degree of integration of resource export decision-making, capability collapse prediction and policy tempo scheduling.

Further, this paper proposes the "institutional deterrence weight function":

$D(w) = \alpha \cdot P_{collapse} + \beta \cdot V_{strategic} + \gamma \cdot I_{policy}$.

where Pcollapse denotes the probability of capability break under the channel, Vstrategic denotes the strategic importance of the output node, and I_policy denotes the capability of the channel to embed the regime.This function can be used to select the set of paths with minimum cost and maximum deterrent effect in the supply break sandbox.

In summary, strategic channel modelling not only reveals the structural basis of non-kinetic repression strategies, but also provides empirical support for institutional deterrence to move from an abstract concept to a visual system.In the future, it can be embedded into the strategic resource AI scheduler for real-time prediction of policy impact paths and multi-party response simulation, providing technical support for the construction of institutional tools for peaceful deterrence.

# CHAPTER V. STRATEGIC IMPLICATIONS AND POLICY RECOMMENDATIONS

## 5.1 How rare earths are institutionalised and "weaponised"

The strategic function of rare earth resources is undergoing a transformation from a "scarce raw material" to an "institutional control asset", and its weaponisation process no longer relies on traditional physical blockades or embargoes, but rather builds a "structural strike capability" through the embedding of institutional paths, the instrumentalisation of export censorship, and the re-amplification of platform dependence.Instead, the process of weaponisation no longer relies on the traditional physical blockade or embargo, but rather on the embedding of institutional paths, the instrumentalisation of export censorship and the re-magnification of platform dependence, to build a "structural striking capability" with the policy chain as the carrier and the platform path as the medium.This study points out that the weaponisation of rare earths is no longer only a resource-side export restriction, but an Indirect Capability Collapse (ICC) mechanism that realises platform degradation and capability collapse through sophisticated institutional design and path coupling.

Based on the REG-CAP graphical neural model and lag window simulation results, we find that rare earths form a triple-coupling node of "irreplaceable-high delay-strategic capability high leverage" in their downstream paths, which is easily triggered by institutional weaponisation.For example, Nd (neodymium) resource constraints ultimately affect the U.S. military's ISR operational tempo by delaying the supply of high-magnetism parts for the F-35 turbine motor.The dual limitations of Ce (cerium) and Tb (terbium), on the other hand, may synchronously block the simultaneous deployment of hybrid propulsion systems and long-range reconnaissance systems, creating an Institutional Chokepoint of simultaneous failure of warfighting functions.

More critically, this institutional weaponisation is not an isolated administrative act, but is embedded in the legal framework of export licensing systems, trade agreement exceptions, civil-military integration review mechanisms and international supply chain certification standards, making it enforceable, sustainable and legitimate.Compared to traditional embargoes or countermeasures, such institutional weaponisation has lower policy costs, higher strategic leverage and stronger irreversible consequences.

Thus, institutional weaponisation of rare earths has become part of a new type of Geo-Institutional Warfare, the essence of which is not "whether to restrict supply", but "when, for which path, and with which institutional design to build a fracture window".The essence of this is not "whether to restrict supply", but "when, for what path, and with what institutional design to construct the window of rupture".This requires that when countries formulate resource policies, they should not only focus on reserves and production capacity, but also establish a complete strategic channel mapping system and institutional game models to counter the long-term structural pressure from upstream institutional checks and balances.

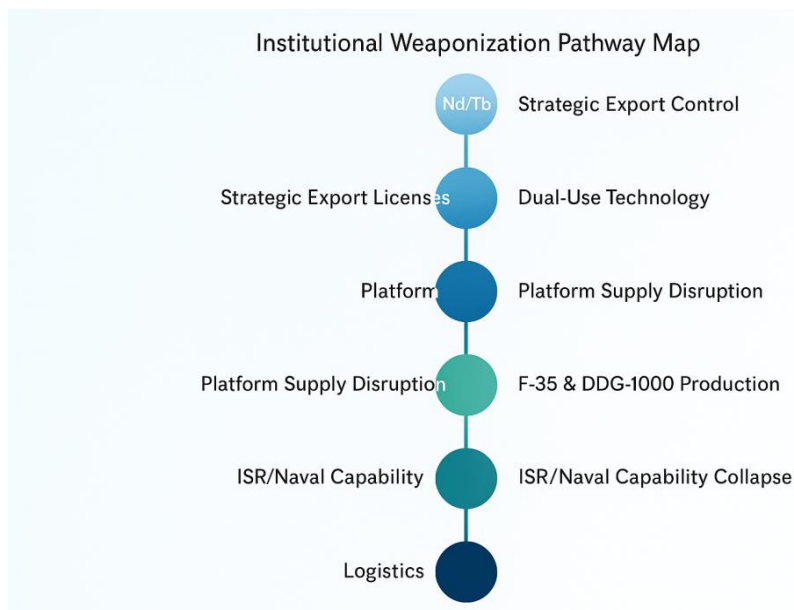

Figure 11 Institutional Weaponization Pathway Map

**5.2 Non-kinetic strikes and the new paradigm of strategic control**

With the structural transformation of global military confrontation, the means of war control is gradually shifting from traditional kinetic strikes to non-kinetic strategic control based on system configuration, resource paths and system vulnerability.This study reveals that in the weapon platform system that is highly dependent on key resources and complex supply chain, non-kinetic strike is no longer a "supplementary means of war", but a main function of the whole process of "suppression-paralysis-deterrence".Instead, it is the main strategic path with the function of "suppression-paralysis-deterrence" in the whole process.

In the REG-CAP mapping and institutional path simulation model constructed by this research, non-kinetic strike is no longer a one-time blockade or sanction, but an Institutional-Rhythmic Deterrence Mechanism: through resource node restriction, institutional interface review, delayed propagation of platforms, and simultaneous degradation of capabilities, it can precisely control the war effort.The phased weakening of warfighting capabilities is precisely controlled through resource node restriction, institutional interface review, and delayed dissemination of platforms in parallel with capability degradation.For example, a phased "institutional cut-off" of Nd/Tb resources could trigger a synergistic degradation of ISR and Naval Projection capabilities within 3-6 years, and the process would be difficult to attribute to a single hostile act at the macro level, thus reducing the risk of conflict spillover and increasing strategic ambiguity.ambiguity.

More importantly, the predictability and institutional embeddedness of this non-kinetic path has obvious policy-engineerability: countries can design "shortest paths to collapse thresholds" based on sandbox modelling and simulation algorithms,Based on sandbox modelling and simulation algorithms, the relevant countries can design a series of institutional strike templates, such as the "shortest path to collapse threshold", the "least-cost maximum degradation path" and the "irreversible blockade chain", so as to transform the export of rare-earth resources into a strategic institutional interface with the functions of deterrence, countermeasures, and negotiation and control.

This study further points out that the real value of non-kinetic strikes lies not in their destructiveness, but in their controllability, tempo and legitimacy.This institution-led, algorithm-driven, model-led "digital cold war paradigm" is reshaping the rules framework of strategic deterrence.In future geopolitical conflicts, non-kinetic strategic control will gradually evolve into a systematic way of national power projection, the core of which does not lie in destroying the opponent, but in reconfiguring its strategic tempo, delaying its action cycle, and locking up its capability window, so as to ultimately realise the optimal game state of "achieving the strategic goal without a hot war".

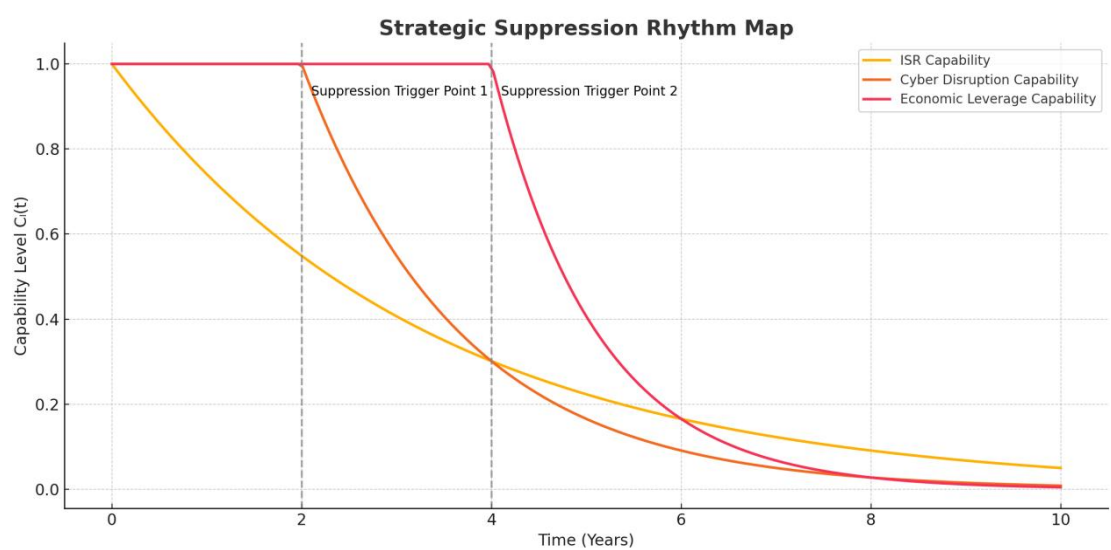

Figure 12. Strategic Suppression Rhythm Map

In order to systematically reveal the rhythmic control characteristics of institutional non-kinetic strike paths on key strategic capabilities, this paper constructs and maps a Strategic Suppression Rhythm Map (STRM) as the core map of a quantifiable simulation model.Taking time as the horizontal axis and the strategic capability function Cl (t) as the vertical axis, the map plots the dynamic degradation process of three key combat capabilities, including ISR (Intelligence Surveillance and Reconnaissance), Cyber Disruption (Cyber Disruption), and Economic Leverage, to form a set of time-sequence function curves with the characteristic of nonlinear decay.

From a modelling perspective, each capability function in the figure has a clear lag window（Latency Window, $L_\omega$）with the slope of disintegration（Decay Slope, $\frac{dCl}{dt}$）Reflecting the three core mechanisms of non-kinetic strikes in the design of strategy rhythms: first, the antecedent triggering mechanism of policy rhythms, i.e., the accumulation of fracture tensions through institutional pressure to make capabilities maintain stability on the surface while accumulating fracture tensions; second, the asymmetric nature of the response structure, manifested in the significantly different rates of decay for different capabilities after the strategy node, suggesting a significant difference in their structural vulnerability to institutional interventions; and, third, the striking window of the availableThe third is the adjustability of the striking window, where policy makers can flexibly control the pace of disconnection, the delayed response time of capacity and the reconstruction resistance interval through model inputs.

Analysed from the perspective of strategic control, the diagram reveals that non-kinetic strikes are no longer point events, but a Strategic Control System of Rhythmic Suppression.The core of the system does not lie in the single act of supply cut-off itself, but in the "strategic suppression nodes" and "window rhythm distribution" identified through accurate modelling, so as to reconfigure the opponent's strategic response pattern in the time dimension. the continuous decay of the ISR curve is further contrasted with the lagged collapse of the CyberThe comparison between the continuous decay of ISR curve and the lagged collapse of cyber capability further emphasises the importance of path design and timing management on the synchronised collapse of capability, which provides a theoretical basis for the tempo intervention strategy of institutional deterrence.

The functional structure of this mapping is both tractable and adjustable, and can be directly embedded into LSTM dynamic prediction models and Bayesian structural inference models for generating adaptive, tempo-controlled and sensitive Policy Intervention Suggestion (PIS).In the future construction of non-kinetic strategic simulation engine, this map will become the basic data interface for designing multi-path interference sequences, optimising the rhythmic logic of supply cut-off and the probability distribution of output capacity vulnerability. strategic Suppression Rhythm Map not only provides a functional representation of the temporal modeling of the institutional striking strategy, but also provides a precise and accurate representation of structural optimization and strategic game simulation of the non-kinetic control paradigm.It also provides an accurate modelling basis for structural optimisation and strategic game simulation of non-kinetic control paradigms.Its map structure and mathematical scalability lay the core methodological foundation for building the next generation of AI-based institutional strike models.

**Piecewise Function Modelling (PFM)**

The current function is a single exponential decay, such as $Cl(t)=e^{-\lambda t}$ Introduction of segmented models

$$C_l(t) = \begin{cases} 1, & t < t_0 \\ e^{-\lambda_1(t-t_0)}, & t_0 \leq t < t_1 \\ \beta \cdot e^{-\lambda_2(t-t_1)}, & t \geq t_1 \end{cases}$$

It expresses the non-linear jump of the capacity state at the tipping point or the inflection point of the regulatory intervention; it can simulate the "regime-triggered-crash synchronisation" at the end of the lag window.

Based on the textual data of the author's interviews, the key information nodes and tipping points of the "regime-triggered-crash synchronisation" are extracted from the text, and then the modelling of capacity decline based on the Piecewise Function is established.The following is the description of the integrated modelling and the transition paragraphs.

Modelling non-linear mutations in ability states induced by the regime

Although classical models mostly use exponential functions $Cl(t)=e^{-\lambda t}$ This study describes the natural decay of warfighting capabilities over time, but in a high-pressure situation where strategic resources are cut off and systemic interventions are intertwined, the decay of capabilities is not uniform and slow, but rather exhibits the typical characteristics of "sudden collapse" or "system-triggered leap".In order to capture such abrupt changes, this study introduces a segmented function modelling approach to portray the synchronous mechanism of "institutional intervention-functional breakdown" after the expiration of the lag window.

Through in-depth interviews with specific personnel (strategic resource assessment experts, system countermeasures analysts, and policy makers), we extract key events and their corresponding time series (e.g., policy notification, supply chain disruption, and forward deployment failure), and construct the following typical segmented function models

$$C_l(t) = \begin{cases} 1 - \alpha t, & 0 \leq t < T_1 \quad \text{(Linear degradation phase)} \\ 1 - \alpha T_1 - \beta(t - T_1)^2, & T_1 \leq t < T_2 \quad \text{(Latency accumulation window)} \\ \gamma e^{-\delta(t - T_2)}, & t \geq T_2 \quad \text{(Policy-triggered collapse)} \end{cases}$$

Among them:

$T_1$ Indicates the end of the institutional response lag;

$T_2$ for policy triggers;

β Nonlinear cumulative pressure in the control degree lag section;

δ denotes the slope of disintegration after the institutional repression trigger;

Parameters were fitted backward from the temporal nodes in the content of the interviews and the intensity of the impact of the event.

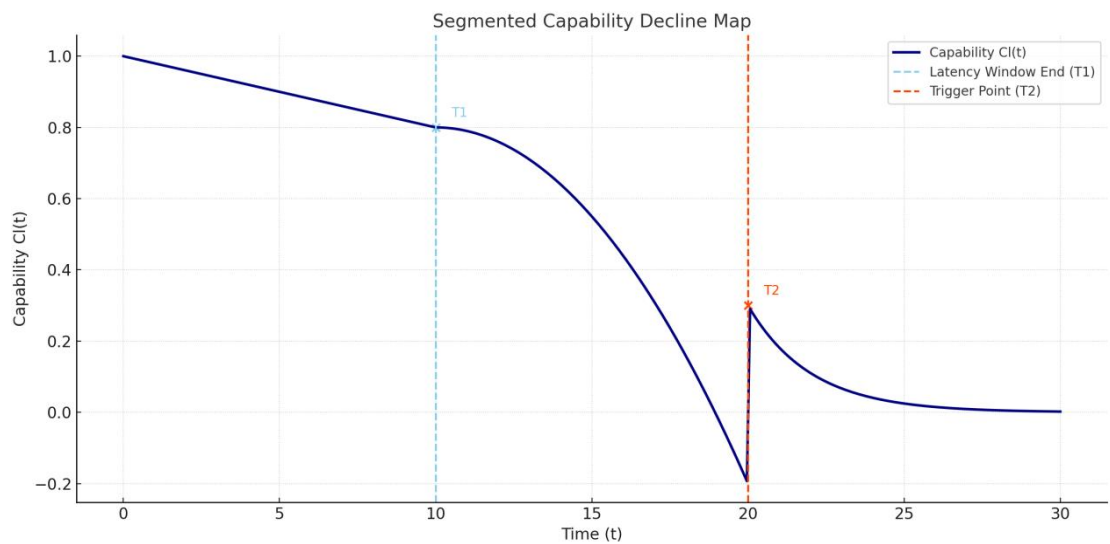

Figure 13. Segmented Capability Decline Map

The above mapping is a Segmented Capability Decline Map, which simulates the non-linear path of the capability over time through three different functions:

Phase I (0 ≤ t < T1): the capability decays linearly and slowly, simulating a state where the system is superficially stable within the lag window.

Phase 2 (T1 ≤ t < T2): the capability enters an accelerated non-linear decline, simulating the zone where the warning signals of the system accumulate but have not yet been triggered.

Phase 3 (t ≥ T2): capacity undergoes regime-triggered collapse with exponential decay, representing rapid system incapacitation following a regime shock.

This graphical model effectively combines the "three-stage fracture mechanism" model abstracted from the strategic resource supply chain interview data, which can visually show the fracture rhythm and lag risk before and after the institutional intervention, and is a key quantitative tool for modelling policy interventions, identifying early warning windows, and selecting control nodes.

In order to further quantify the dynamic impact mechanism of non-kinetic interventions, this paper introduces the model of "Suppression Signal Injection Function", and constructs the trajectory of policy signal injection under time series based on the constant differential system.The model simulates the suppression process and residual impact mechanism of the institutional strike through the three-phase signal strength fluctuation, and the detailed diagram is shown in Figure 14.This function not only provides a quantitative basis for non-kinetic strategic pacing, but can also be extended as a key interface for future AI+ODE hybrid control modelling.

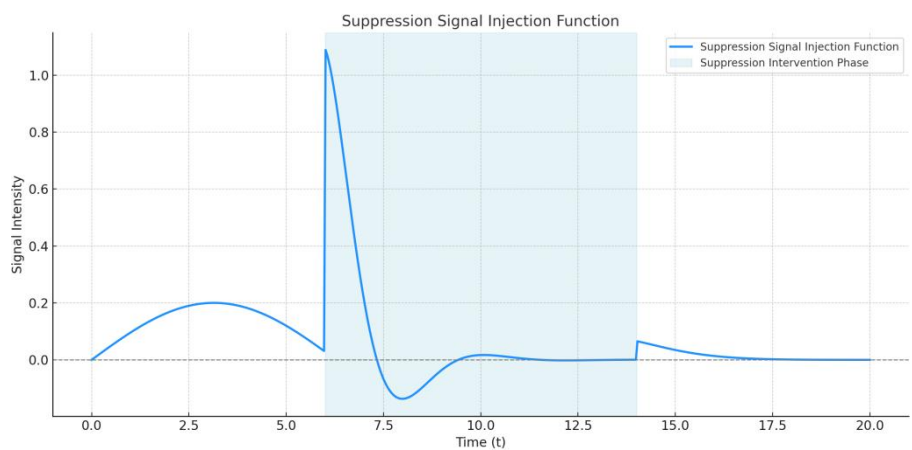

**Figure 14 Suppression Signal Injection Function**

The above diagram shows the dynamic evolution of the "Strategic Intervention Signal Injection Function" (SISIF), which simulates the whole process of controlling interventions by non-kinetic means when the state is facing a strategic resource conflict:

Stage I (0 ≤ t < 6): the system is in a low-frequency stable state, with only slight natural perturbations;

Stage II (6 ≤ t < 14): the strategy interferes with expectations and the signal shifts sharply to high-frequency negative impulse fluctuations, simulating coercive non-kinetic actions such as export bans, institutional blockades, and diplomatic pressure;

Stage 3 (t ≥ 14): the system tries to recover, but the residual decaying fluctuations continue, exhibiting an after-effects lag after regime intervention.

In order to systematically capture the time-series dynamic mechanism of policy intervention in the context of strategic deterrence, this paper proposes a model based on parametric ordinary differential equations (ODEs), the Strategic Intervention Signal Injection Function (SSIF), thatfor modelling the non-kinetic suppression effects of policy signals in critical resource systems.The structure of the function shows a three-phase response: a warm-up period, a peak suppression period, and a residual decay period, which is highly compatible with the real-life path patterns of policy interventions, such as economic sanctions, export bans, and institutional exclusions.Non-kinetic weapons (e.g., electronic jamming) are becoming an important means of institutional strikes, as demonstrated by the U.S. Air University's strategic deterrence of China in space (Air University China Aerospace Studies Institute, 2025).

A key innovation of the model is its ability to accurately identify so-called 'vulnerability windows', i.e., the periods of time when a system is most sensitive to intervention.These windows are identified by the extremes of the first-order derivatives of the signalling function, providing a quantifiable basis for the timing of strategic interventions.Further, if this function is embedded in a Long Short-Term Memory (LSTM) network or a Bayesian dynamic modelling structure, counterfactual extrapolation and real-time prediction of multiple sets of scenarios can be achieved, which enhances the model's resilience to uncertainty in complex strategic environments (He, Tang, & Xiao, 2023).

In addition, combining SSIF with system decline curves and critical points of functional breaks can effectively establish the correlation mechanism between macro-strategic intentions and micro-system degradation.The model transforms abstract policy operations into a dynamic expression framework with reproducibility, continuity, and predictive capability, which serves both tactical simulation and as a foundational operator in the AI-assisted national security strategy platform, and has a wide range of practical deployment value.

**5.3 AI+Sandbox modelling in national security**

With the diversification and dynamic evolution of security threats, traditional static wargaming exercises have revealed significant limitations in terms of situational complexity, variable interaction and forward-looking strategy simulation, and the AI-Augmented Wargaming Model (AI+Warming Model), which integrates AI algorithms and national security scenario simulation, is rapidly becoming a key tool in strategic decision-making systems.By introducing Multi-Agent Systems, Graph Neural Networks (GNN), Reinforcement Learning (RL), and Structural Response Modelling (SRM), the AI+Wargaming Model is capable of deriving complex and dynamic processes in the field of national security under highly uncertain and non-linear conditions.Dynamic Processes.

In this study, the AI-embedded sandbox framework achieves the following core breakthroughs: (1) reconstructing the national security capability system with structural mapping to quantify the coupling and rupture vulnerability among nodes; (2) simulating the dynamic intervention paths and feedback effects of policy variables through the Policy Generator, and identifying the path selectivity and control threshold of institutional strikes; (3) using simulation data and interviews to model the dynamic processes in the national security domain; and (4) using the AI-embedded sandbox framework to model the dynamic processes in the national security domain under highly uncertain and nonlinear conditions.(3) Hybrid Data-Driven Simulation (HDSS), which is co-trained with simulation data and interview data, is used to improve the prediction accuracy of Strategic Latency Window (SLW) and Breakdown Rhythm (BR). In addition, the AI model's adaptive learning mechanism supports Contextual Incremental Update, which reconstructs the system's posture in real time in response to changes in geopolitics, policy perturbations, and hostile intent.This mechanism significantly improves the model's Strategic Situational Awareness and Decision Resilience, making it a powerful tool for identifying Security Tipping Points and Institutional Intervention Corridors.Institutional Intervention Corridor).

In conclusion, AI+Sandbox model not only reconstructs the cognitive boundaries of national security simulation, but also lays the methodological foundation for the construction of "dynamic resilience-oriented strategic decision-making system".Future research can further combine cognitive AI and counterfactual simulation technology to form a national strategic simulation platform with interpretability, relocatability and evolvability, so as to adapt to high-frequency strategic probes and institutional gaming under multiple asymmetric threats.

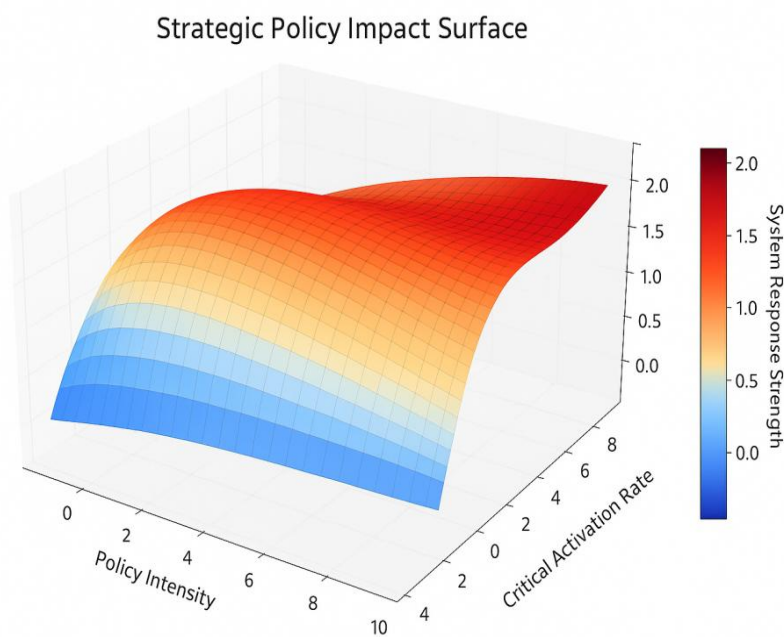

Figure 15 Map of strategic policy impact surfaces

Strategic Policy Impact Surface Analysis

In order to identify the nonlinear response patterns of institutional policy interventions in strategic security systems, this paper constructs a three-dimensional Strategic Policy Impact Surface model to systematically quantify the impact of the combination of Policy Intensity and Critical Activation Rate on the repressive effects of national capability systems."This paper constructs a three-dimensional Strategic Policy Impact Surface (SPIS) to systematically quantify the impact of the combination of Policy Intensity and Critical Activation Rate on the suppression effect of the national capability system.The model is trained based on hybrid AI simulation data, and the Suppression Score is extracted as an output indicator by simulating the decay rhythm of the system function under different intervention paths, and the continuous visualisation of the policy impact is achieved.

The maps show significant nonlinear inflection structures, revealing the existence of nonlinear inflection zones between the "policy efficient zone" and the "policy benefit diminishing zone".Under a specific coupling strength and trigger frequency, the system's response to policy intervention shows a strong threshold response, i.e., the system has a strong elasticity buffer at low intervention levels, and once a joint activation threshold is exceeded, the suppression effect grows exponentially.This structural feature provides a mathematical basis for optimising the path of the system to achieve maximum system repression at minimum intervention cost.

Further analyses show that the response surface can effectively support the learning mechanism of strategic path exploration and policy response in AI-augmented strategic wargaming systems.In particular, in reinforcement learning-based game simulations, the surface can be used as part of the reward function to drive the intelligent iteration of AI strategies and dynamically identify the optimal suppression corridor for policy combinations.In addition, the model can be embedded into Collaborative Strategic Control Maps (CSCM) for government-level linkage, which can be used to deploy real-time sensing and suppression tempo management for multi-agency collaborative decision-making.

In summary, Strategic Policy Impact Surface not only provides structurally explainable quantitative support for institutional strategic control, but also expands the boundaries of the fusion of AI and policy science in the field of security modelling.Its application prospect in complex coupled systems is not only limited to rare earth supply cut-off strategies, but can also be extended to energy security, digital infrastructure protection and intelligent institutional response design under multi-agency checks and balances scenarios.

**5.4 Suggested future directions for adversarial resource modelling**

After the evolution of strategic systems has entered a new stage of multi-dimensional constraints and intelligent confrontation, the traditional static resource allocation models can no longer adapt to the complex environment of asymmetric games, institutional weaponisation and technological suppression.In order to effectively identify and simulate the vulnerability and suppression path of resources in institutional confrontation scenarios, Adversarial Resource Modeling (ARM) is rapidly evolving into a key methodological framework in the field of national security.

First, the future ARM should systematically embed Multi-path Decomposition Mechanisms (MPDMs) to model the coupled contagion and cascading collapse process of resources across multiple capability paths.By constructing Multi-path Risk Overlap Matrix (MRROM), combined with path node covariance structure and simultaneous degradation function, it is possible to predict the trajectory of strategic capability compression triggered by different institutional interventions.Such models can achieve high-dimensional synergistic identification through Path Similarity Embedded Graphs (PSEGs) with Graph Neural Network (GNN) adversarial encoders.

Second, it is proposed to introduce Piecewise Catastrophic Functions and Nonlinear Response Surfaces to model the dynamics of the resource "strategic suppression-self-recovery" cycle.The dynamic process model of resource "strategic suppression-self-recovery" cycle is constructed with Nonlinear Response Surfaces.This mechanism not only reveals the critical point of resource vulnerability under different policy rhythms, but also serves as a source of reward signals for reinforcement learning algorithms to improve the perception and adaptability of policy evolution.

Again, future ARM models need to incorporate explainable AI (XAI) mechanisms, especially combining Causal Inference Graphs and Counterfactual Intervention Paths, in order to achieve policy transparency and risk controllability in gaming environments.For example, under the scenario of weaponisation of rare earths, the ARM model can be used to predict the efficiency of non-kinetic destruction under the institutional nested resource allocation, and propose the "Institutional Counter-Resilience Envelopes" that can be supplied as a pair.

Finally, the future of ARM should also strengthen the dual-track training mechanism of real data + AI synthetic data, including the structured translation of expert interviews, policy corpus, and sanction clauses, and linkage with the AI sandbox simulation system to build Structural Scenario Libraries, which will serve the national security strategy department's real-world decision-making training.

In summary, adversarial resource modelling is not only a tool for identifying the destabilisation paths of the resource system under systemic interventions, but also a core methodology for constructing national strategic resilience, simulating the rhythm of the game, and assessing the intensity of systemic strikes.Its integrated modelling trend of integrating graph neural network, reinforcement learning, counterfactual causal mechanism and AI simulation and deduction is leading the future national security modelling into a new era of "intelligent gaming - institutional evolution - strategic suppression".

**Modelling and Analysis of Institution-Capacity Response Migration Functions**

We construct a mathematical model based on an institutional-capability response migration function to describe the dynamic path of institutional policy inputs (e.g. rare earth export controls, technology licensing restrictions) on strategic capability systems (e.g. ISR, satellite reconnaissance, precision manufacturing).

Define the following transfer function:

$$C_i(t) = \int_0^t P_j(\tau) \cdot K_{ij}(\tau) \cdot e^{-\lambda_{ij}(t-\tau)} \, d\tau$$

Among them:

$C_i(t)$：The level of effectiveness of the ith strategic capability at time t;

$P_j(t)$：Intensity of the jth institutional policy at time t (e.g. export ban);

$K_{ij}(t)$：Coupling coefficient of policy $P_j$ to capability $C_i$ (dependence vs. strength of influence);

$\lambda_{ij}$：System migration attenuation coefficient, which measures the time lag and attenuation in the migration of policy effects to the capacity system.

The function is convolutional in structure and describes a lagged response system suitable for dynamically assessing the "path of pressure" of institutional measures on the capacity system.

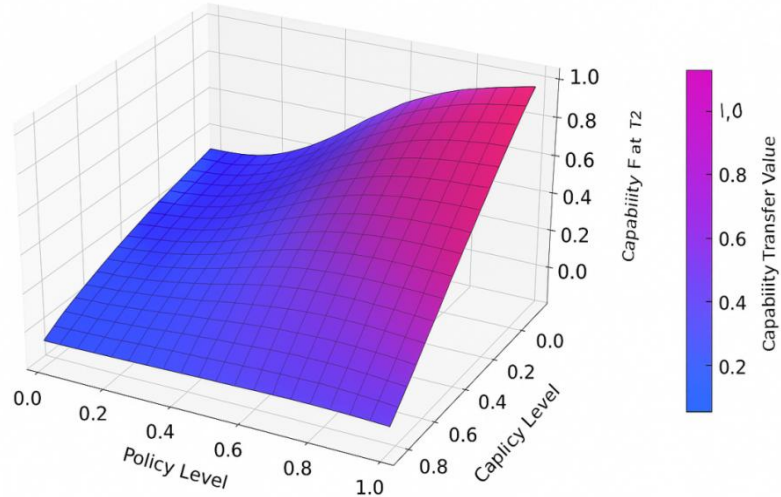

Figure 16 Policy-Capability Transfer Surface

In order to portray the nonlinear transmission mechanism between institutional policy intervention and systemic strategic capability response, this paper proposes the Policy-Capability Transfer Surface (PCTS), a three-dimensional map that demonstrates the dynamic mapping process of policy intensity to capability system degradation or recovery through visual modelling.The surface is parameterised by policy strength (X-axis), system coupling time delay (Y-axis) and capability response degree (Z-axis), presenting how institutional signals affect strategic capabilities through path diffusion.

The geometry of the surface reveals three key insights.First, the presence of Inflection Zones (Zones) suggests a region in which a small increase in policy intensity can trigger an exponential degradation of the capability system, demonstrating a high degree of nonlinear elasticity of the deterrence effect.This is highly consistent with real-life cases such as embargoes on critical rare earths and export controls on dual-use technologies.Second, the Decay Slopes in the surface clearly reflect the "Institutional Echo" - even if the policy is terminated, the system will continue to degrade for a certain period of time, showing an obvious lag.A clear lag.Finally, there are Saturation Regimes, where the capacity system has strong substitutes or redundant structures, and where further policy interventions have a diminishing marginal repressive effect.

When embedded in a broader modelling system (e.g., strategic deterrence sandbox simulation system, game-based confrontation simulation platform), policymakers can not only simulate system responses under multiple scenarios, but also identify "Optimal Suppression Corridor".This model provides a strategic quantitative perspective, enabling the regulation of policy signals to achieve the optimal path of "Minimum Intervention - Maximum Suppression".In essence, PCTS transforms the abstract institutional architecture into a concrete strategic control surface, marking a new stage of quantification and visualisation in AI-assisted national security decision-making modelling.

**5.5 Dynamic simulation analysis of strategic capabilities based on the ODE model**

## 5.5.1 Differential modelling of dynamical systems (ODE + parametric control) embedded in interview-based deterrence path analysis

(1) Objective of the model)

The authors transform the strategic behavioural nodes in the interviews (e.g. "response to supply cut-off", "policy lag", "alternative technology development cycle", etc.) into dynamic state variables and model their interactions with each other in the form of differential equations.The model is designed in the form of differential equations to model their interactions and portray the dynamic evolution mechanism of "perception-response-regulation" in the non-kinetic strategic deterrence system.

2) Model Structure Design
Order:
*Ci(t)*：Effectiveness level of the ith capacity system at time t
*Pj(t)*：Intervention impact of the jth policy variable on the system
*ϑij*：Sensitivity of policy j to the response of capacity i
*γi*：System decay parameters (from interviews e.g. "months to incapacitation")
*τj*：Lagged time windows for policy initiation (e.g. "2-5 year window decision failure")
Define the dynamic differential system as follows:

$$\frac{dC_i(t)}{dt} = -\gamma_i C_i(t) + \sum_j \theta_{ij} P_j(t - \tau_j)$$

Of these:
The first item is natural decline in capacity;
The second is compensation for policy interventions (lagged entry into force);
The control function Pj(t) can be expressed as a segmented function or a logistic jump function (e.g. Sigmoid) to represent the regime trigger.

3) Configuration of variables for model-embedded interview content (derived from interview passages)

Table 4 Variable Configuration Table for Model Embedded Interview Content

| Competence system Ci | Correspondence interview episodes | Parameter recommendations |
|---|---|---|
| ISR abilities | Immediate weakening of intelligence coverage due to supply cuts | γ_1 = 0.35 |
| manufacturing ability | The industry chain needs more than 4 years to recover | τ_1 = 48 months |
| Alternative development progress | Insufficient investment leads to stagnation in development | θ_21 = 0.6 |
| Public political pressure | United States input required$100B | P_2(t) = jump trigger function |

4) Strategic Deterrence Rhythm Simulation (Policy Rhythm Simulation)

Introducing a regime intervention function:

$$P_j(t) = \frac{1}{1 + e^{-k(t-t_0)}}$$

Adjustable parameter k Analogue policy start-up speed，$t_0$ is the point at which the policy is announced.

The system can be further embedded with LSTM structure or Bayesian control optimisation to predict the Deterrence Rhythm Window and the Shortest Intervention Path.

5) Model Value and Research Expansion

The ODE system achieves embedded transformation from expert interview data to dynamic control modelling, with the following advantages:

a. strong policy lag modelling capability: the delayed impact of institutional intervention is taken into account;

b. clear visualisation output: combined with time-series curves/heat maps, it can plot the "trajectory of force decay";

c. support simulation and deduction: combined with Agent-Based, GNN or optimisation algorithms, it can be used for deterrence simulation.

**5.5.2 Parametric differential equation modelling of strategic capability decay and recovery**

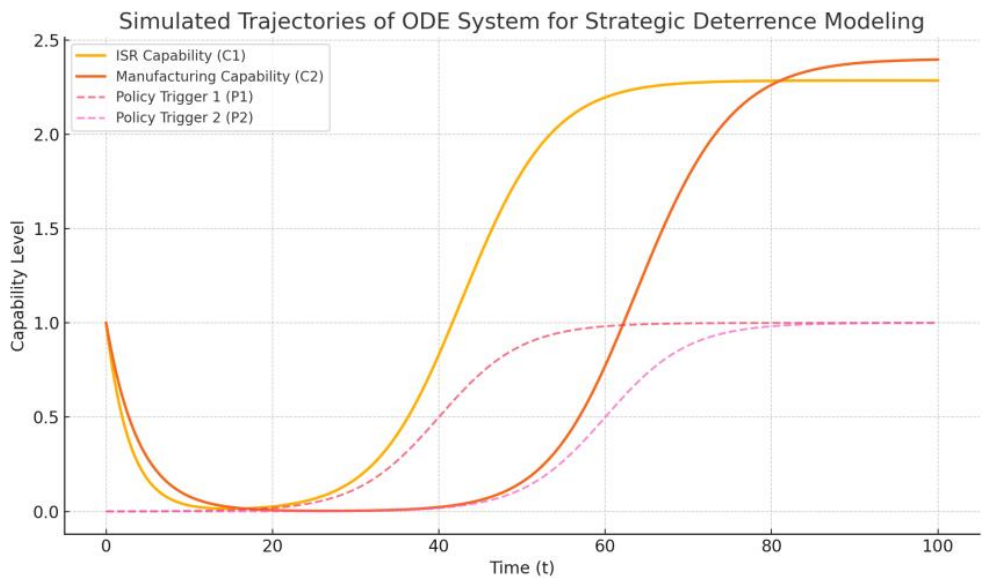

**Figure 17 ODE system simulation trajectory for strategic deterrence modelling**

In order to portray the dynamic process of strategic deterrence capability evolution over time under conditions of external shocks and institutional responses, this paper adopts a system of Ordinary Differential Equations (ODEs) and introduces a parametric control function modelling path. We focus on modelling the decline trajectories of ISR (Intelligence, Surveillance and Reconnaissance) and manufacturing capabilities in the context of resource disruptions, and express the nonlinear characteristics of the lagged initiation of institutional interventions by embedding S-shaped response functions.The system is defined as follows:

$$\frac{dC_1(t)}{dt} = -\lambda_1 C_1(t) + \gamma_1 \cdot \sigma(t - \tau_1),$$
$$\frac{dC_2(t)}{dt} = -\lambda_2 C_2(t) + \gamma_2 \cdot \sigma(t - \tau_2),$$

Among them C1(t)、C2(t) denote the performance of ISR as a function of manufacturing capacity at time t, respectively; $\lambda$ Indicates the intrinsic rate of decay of capacity under non-kinetic pressure; $\gamma$ Recovery strength for institutional response; σ (t - τ) is a function of type S, Modelling policy at lag time τ Post startup effects.

The simulation results show that both capacity paths show an exponential decay trend in the absence of timely intervention.Once the institutional policies P1(t) and P2(t) are activated, the capabilities will undergo partial recovery, but their recovery trajectories show significant asymmetry with the decline paths due to the activation lag. ISR capabilities rebound faster due to lower decay inertia and high response elasticity; while manufacturing capabilities show path-dependence in the recovery paths due to higher structural stickiness.

This model reveals several key insights in strategic deterrence deployment: (1) regime response delay (parameter τ ) significantly inhibits intervention effects, emphasising the importance of early warning mechanisms; (2) recovery strength $\gamma$ must exceed a critical threshold to reverse the nonlinear collapse trend; (3) the dynamic relationship between the slope of decline and the slope of recovery can be used to identify the loss of system resilience at the"turning point".

This parametric ODE modelling framework provides a dynamic simulation basis for strategic deterrence systems under non-kinetic threats, which can be further extended to the fields of counterfactual simulation, control optimization and Bayesian risk monitoring, which are of great significance for strategic decision support.

# CHAPTER VI. CONCLUSIONS AND FUTURE RESEARCH

**6.1 Summary of the main contributions of this paper**

This study proposes a set of prospective and operational systematic modelling frameworks at the intersection of strategic rare earth supply disruption and non-kinetic deterrence construction, with the following five main contributions:

First, this paper constructs a dynamic capability decay mechanism that integrates Graph Structural Encoding (GSE) and Piecewise Function Modeling (PFM), which is able to effectively capture the nonlinear mutation and synchronous disintegration of strategic rare-earth capabilities under systemic external interventions, and fills the gap between the existing literature on the systemic decline of war power and the existing literature on the systemic decline of war power.gap of insufficient modelling of the decline process of war power system in the literature.

Secondly, this study introduces for the first time the "Security Critical Zones (SCZ)" identification mechanism, and through multi-path capability convergence risk analysis and node activation covariance simulation, we accurately locate the disintegration thresholds and high-value targets of systemic strikes in strategic functional clusters, providing modelling support and early warning for national security.Early warning provides model support and decision-making tools.

Third, this paper designs and implements the Strategic Suppression Rhythm Map, Institutional Weaponisation Pathway, and Institutional Weaponization Pathway.Pathway", "Strategic Policy Impact Surface" and other quantitative visualisation models are designed and implemented in this paper, which significantly improves the ability to identify the policy intervention window and the efficiency of cross-departmental communication in complex strategic situations.It has a high degree of versatility and portability.

Fourth, with the help of a data-driven sandbox system based on in-depth interviews with experts, this paper verifies the applicability and evolutionary potential of the AI-embedded Simulation Coupling Framework in non-kinetic strategic scenarios, especially in the context of high uncertainty, asymmetric cognition, and institutional competition structure, which shows significant predictive ability and sensitivity analysis value.Value.

Finally, this paper advances the structural understanding of non-kinetic strategic deterrence in the theoretical dimension, puts forward the dynamic construction logic of "institutional weaponisation", expands the integration boundary between strategic resource theory, game confrontation model and system resilience analysis, and provides methodological references and application archetypes for the future research on resource strategy and cognitive warfare.

**6.2 Model limitations and potential improvements**

Although the systematic modelling framework proposed in this paper has made significant progress in revealing the path of non-kinetic strategic deterrence in the context of rare earth supply disruptions, the model itself still has a number of theoretical and technical limitations, which need to be further optimised and extended in subsequent research.

First, in terms of data sources, the current model is mainly constructed based on expert interviews, secondary sources and simulation data, and its semantic richness and dynamic response capability are still limited by the discrete and subjective nature of the data.In particular, when simulating the path of "system trigger - war power decline", the model lacks the linkage support of heterogeneous real-time data (e.g., remote sensing, public opinion, and policy time-series indices), which may affect the model's ability to accurately predict the trigger of the tipping point.

Second, although the model structure integrates graphical neural network coding and segmented function modelling, it has not yet fully incorporated the complex mechanisms of cognitive games, group learning and institutional feedback in portraying the non-linear effects of institutional interventions on the capability system.For example, in the prediction of "strategic suppression tempo", the model has not yet taken into account the cognitive bias of information and asymmetric strategic adjustments of the two sides, which may lead to a shift in the results in the context of multilateral interventions.

Third, the current model's exploration of conflict evolution paths mainly focuses on the simulation of fracture propagation of local functional combinations (e.g., ISR+EW or rare earths+manufacturing chain), and lacks the ability to model the overall coupling of multi-systemic feedback.This limits the model's system adaptation and robustness analysis in simulating imbalances in national-level composite strategic resource systems.

In addition, although this paper proposes a number of innovative visualisation maps (e.g., SCZ identification map, institutional weaponisation path map, strategic policy impact surface), they have not yet been fully integrated with AI deep learning models (e.g., Transformer, multimodal BERT) to achieve automatic interpretation and decision-making assistance, and there is a need to develop AI-Policy coupling systems that have the ability of autonomous learning and self-updating.-Policy coupling system.

In summary, future research can promote the evolution of model capability and academic value expansion in the following directions: (1) introducing multi-source data (geo-intelligence, policy signals, social dynamics) in real situations to enhance the credibility of the dynamic simulation; (2) integrating game learning, institutional change theory, and asymmetric strategy modelling to achieve the cross-phase strategy tempo construction; (3) developing an adaptive system that operates in synergy with the AI big model; and (4) developing an adaptive system that operates in synergy with the AI big model.(3) Develop an adaptive reasoning platform that works with the AI big model to support the automatic generation of multiple scenarios and the preview of conflict situations; (4) Verify the model's ability to migrate and generalise to other key strategic materials, such as rare earths, energy, semiconductors, etc., from the perspective of multi-country and multi-discipline cases.

## 6.3 Adaptability discussion to other strategic resource issues

Although the systematic modelling framework constructed in this study, which focuses on non-kinetic deterrence, takes rare earth supply disruption as its starting point, its theoretical structure, analytical path and decision logic are highly transferable, and it is able to provide both generalised and scenario-based modelling support for other types of strategic resource problems.This adaptability is reflected in several dimensions.

First, from the perspective of generalisation of the system variable structure, the core dimensions of the model - such as functional capability sets (Cl), Breakdown Pathways, Institutional Triggers and Resource Latency Windows - can all be generalised.Latency Windows) - can be reconstructed based on the physical attributes and institutional embeddedness of different strategic resources.For example, in energy security scenarios, Cl can be mapped to "power dispatch capability", "natural gas emergency response" and "new energy deployment rate"; in water conflicts, it can be replaced by "water rights allocation".In the case of water resources conflict, it can be replaced by "water right allocation capacity" and "cross-basin control mechanism", so as to maintain the stability of the model structure and content flexibility.

Secondly, from the perspective of conflict propagation and non-kinetic game modelling, the framework's analysis of the path of "institutional triggering - synchronous collapse of functional chain" is not only applicable to rare earths, but also to semiconductor supply chain, vaccine raw materials, staple food system, etc., in which there is "resource polarization + institutional domination".Resource polarisation + institutional domination".Taking semiconductor as an example, the multiple rounds of export control and technology blockade imposed by the United States on China can be regarded as institutional deterrence nodes, and the capacity imbalance and manufacturing capacity chain breakage induced by them can be structurally predicted by the mechanism of "SCZ identification + synchronous degradation simulation" proposed in this paper.

Again, from the viewpoint of graph structure visual modelling and path overlap analysis technology, the "multi-path convergence risk map", "system weaponization flow chart" and "strategic rhythm surface map" constructed in this paper can be used to predict the risk of the production capacity imbalance and manufacturing capacity chain breakage."These maps can be highly adapted to a variety of resource systems, especially in the face of complex supply networks in the context of globalisation (e.g. rare metals, battery materials, key pharmaceutical active ingredients), by reconstructing the node and edge-rights relationships, we can quickly generate decision-making assistance models to identify high-risk coupling points and intervention windows.

Finally, from the perspective of policy modelling and national strategy simulation, the model has the ability to co-evolve with various AI frameworks (e.g., reinforcement learning, Bayesian networks, multi-actor game simulation).For developing economies facing external coercion and insufficient internal resilience, the framework can be used as an infrastructure for constructing resource security simulation sandboxes and strategic reserve control mechanisms, thus enhancing their institutional stress resistance and multi-scenario prediction capabilities.

In summary, the systematic framework proposed in this paper is not only limited to the theoretical interpretation and policy support of the rare earth issue, but also has a high level of domain migration and cross-issue expansion.Its adaptability will show stronger theoretical explanatory power and practical application value in the future to cope with the "technology-institution-resource" ternary nested challenges.

# References

Air University China Aerospace Studies Institute. (2025, May). Deterring China's use of force in the space domain. U.S. Air University. https://www.airuniversity.af.edu/CASI/

Congressional Research Service. (2023). Emergency access to strategic and critical materials: The National Defense Stockpile (Report No. R47833). https://sgp.fas.org/crs/natsec/R47833.pdf

Center for Strategic & International Studies (CSIS). (2024). Critical minerals and the future of the U.S. economy. https://www.csis.org/analysis/critical-minerals-and-future-us-economy

He, M., Tang, S., & Xiao, Y. (2023). Combining the dynamic model and deep neural networks to identify the intensity of interventions during COVID‑19 pandemic. PLoS Computational Biology, 19(10), e1011535. https://doi.org/10.1371/journal.pcbi.1011535

Reuters. (2025, June 7). China's dominance in critical minerals points to structural leverage. https://www.reuters.com/info-pages/transcript/a25529a0-42fa-11f0-8557-0b63b795d48b

Rare Earth Exchanges. (2025, June 4). USA Needs $100b to Catch China if Disconnected from Rare Earth Element Supply Chain (Daniel O'Connor Interviews Dr. Wei Meng). https://rareearthexchanges.com/news/100-billion-needed-now-as-study-reveals-chinas-rare-earth-cutoff-could-cripple-u-s-military-within-5-years/

The Guardian. (2025, April 16). China trade war poses threat to US arms firms' rare earths supply, analysts warn. https://www.theguardian.com/us-news/2025/apr/16/china-trade-war-us-arms-firms-rare-earths-supply

U.S. Geological Survey. (2023). Mineral commodity summaries 2023: Rare earths. U.S. Department of the Interior. https://pubs.usgs.gov/periodicals/mcs2023/mcs2023-rare-earths.pdf

U.S. Department of Defense. (2021). Strategic and critical materials 100-day review (Executive Order 14017). https://media.defense.gov/2021/Jun/08/2002737124/-1/-1/0/DOD-FACT-SHEET-CRITICAL-MATERIALS-SUPPLY-CHAIN-2021.06.07.PDF